\documentclass[11pt]{amsart}

\usepackage{xspace,euscript}

\newcommand{\eg}{e.g.,\xspace}
\newcommand{\ie}{i.e.,\xspace}

\theoremstyle{plain}
\newtheorem{thm}{Theorem}
\newtheorem{prop}{Proposition}
\newtheorem{lemma}{Lemma}
\newtheorem{cor}{Corollary}
 
\newtheorem{mainthm}{Main Theorem}

\theoremstyle{definition}
\newtheorem{defn}{Definition}
\newtheorem{prob}{Reduced Problem}

\numberwithin{equation}{section}

\newcommand{\cal}{\EuScript}

\newenvironment{pf}{\proof[\proofname]}{\endproof}
\newenvironment{pf*}[1]{\proof[#1]}{\endproof}

\renewcommand{\epsilon}{\varepsilon}
\newcommand{\R}{\mathbb R}
\newcommand{\A}{\mathbb A}
\newcommand{\bs}{\setminus}
\newcommand{\norm}[1]{\left|#1\right|}

\newcommand{\Norm}[1]{\left\|#1\right\|}
\newcommand{\HR}{\mathbb H_{R}^2}
\newcommand{\HC}{\mathbb H_{C}^2}
\newcommand{\Hy}{\mathbb H}
\newcommand{\Div}{\operatorname{div}}

\newcommand{\vp}{\varphi}
\newcommand{\D}{\partial}
\newcommand{\dist}{\operatorname{dist}}
\newcommand{\compose}{\operatorname{\scriptstyle\circ}}
\renewcommand{\S}{\cal S}
\newcommand{\B}{\cal B}
\renewcommand{\H}{\cal H}
\newcommand{\Hr}{\cal H_R}
\newcommand{\ub}{\bar u}
\newcommand{\vb}{\bar v}
\newcommand{\Qb}{\bar Q}
\newcommand{\Tb}{\bar T}

\newcommand{\inner}[2]{\langle#1,#2\rangle}

\newcommand{\Ric}{\operatorname{Ric}}
\newcommand{\tr}{\operatorname{tr}}
\newcommand{\Lie}{\cal L}
\newcommand{\alphat}{\tilde\alpha}
\newcommand{\betat}{\tilde\beta}
\newcommand{\h}{{\tilde h}}
\newcommand{\vpt}{\tilde\varphi}
\newcommand{\vt}{\tilde v}
\newcommand{\ut}{\tilde u}
\newcommand{\Sigmat}{\widetilde\Sigma}
\newcommand{\Omegat}{\widetilde\Omega}
\newcommand{\uo}{u_0}
\newcommand{\ubh}{\bar u_0}
\newcommand{\Uo}{U_0}
\newcommand{\Uto}{\widetilde U_0}

\newcommand{\Mbar}{\bar M}
\newcommand{\gbar}{\bar g}
\newcommand{\kbar}{\bar k}
\newcommand{\Rbar}{\bar R}
\newcommand{\Ebar}{\bar E}
\newcommand{\Bbar}{\bar B}
\newcommand{\epsilonbar}{\bar\epsilon}
\newcommand{\nablabar}{\overline\nabla}
\newcommand{\cast}{\star}
\newcommand{\uh}{\hat u}
\newcommand{\vh}{\hat v}
\newcommand{\vph}{\hat\varphi}
\newcommand{\Ball}{B}
\newcommand{\Bh}{{\mathbb B}}

\newcommand{\aut}[1]{{\sc #1}}
\newcommand{\tit}[1]{{\em #1\/}}
\newcommand{\vol}[1]{{\bf \underline{#1}}}
\newcommand{\yr}[1]{(#1)}
\newcommand{\pp}[2]{#1--#2}

\begin{document}

\title[Einstein/Maxwell Equations]{$N$-Black hole stationary and axially symmetric
    solutions of the Einstein/Maxwell equations}
\author{Gilbert Weinstein}
\address{Department of Mathematics\\
    University of Alabama at Birmingham\\
    Birmingham, Alabama 35205}
\email{weinstei@math.uab.edu}
\thanks{This research was supported in part by NSF Grant~DMS-9404523.\\
    The author would like to express his thanks for the support
    and hospitality of the
    Erwin Schr\"odinger Institute where part of this work was completed.
    }

\begin{abstract}
The Einstein/Maxwell equations reduce in the
stationary and axially symmetric
case to a harmonic map with prescribed singularities
$\vp\colon\R^3\bs\Sigma\to\HC$, where $\Sigma$ is a subset of the axis of
symmetry, and $\HC$ is the complex hyperbolic plane.
Motivated by this problem, we prove the existence and uniqueness of harmonic
maps with prescribed singularities $\vp\colon\R^n\bs\Sigma\to\Hy$, where
$\Sigma$ is a submanifold of $\R^n$ of co-dimension
$\geq 2$, and $\Hy$ is a classical Riemannian
globally symmetric space of noncompact type and rank one.  This
result, when applied to the black hole problem, yields solutions which
can be interpreted as equilibrium configurations of
multiple co-axially rotating charged
black holes held apart by singular struts.
\end{abstract}

\maketitle

\section{Introduction}

Let $(M,g)$ be a four-dimensional Lorentzian manifold, and let
$F$ be a two-form on $M$.
Consider the Einstein/Maxwell field equations:
\begin{gather}
    \label{eq:einstein}
    \Ric_g -\frac12 R_g \, g = 2 \, T_F\\
    \label{eq:maxwell1}
    F = dA\\
    \label{eq:maxwell2}
    d{\ast F}  = 0,
\end{gather}
where $\Ric_g$ is the
Ricci curvature tensor of the metric $g$, $R_g$ its scalar curvature,
and $T_F$ is the energy-momentum-stress tensor of
the electromagnetic field $F$:
\begin{equation}
    \label{eq:energy:momentum}
    \begin{split}
    T_F(X,Y) &= \frac12 \bigl( i_X F \cdot i_Y F +
    i_X {\ast F} \cdot i_Y {\ast F} \bigr) \\
    &= i_X F \cdot i_Y F - \frac12
    \norm{F}^2  g(X,Y),
    \end{split}
\end{equation}
Here, $i_X$ denotes inner multiplication by the vector $X$,
$\ast$ is the Hodge star operator, $\sigma\cdot\tau$
denotes the inner product, and $\norm{\sigma}^2$ the norm of $k$-forms.
These are given in local coordinates by:
\begin{gather}
    \label{eq:star}
    \ast \sigma_{\mu_1 \dots \mu_{4-k}}
    = \frac{1}{k!}\sigma^{\nu_1\dots \nu_k}
    \epsilon_{\nu_1 \dots \nu_k \mu_1 \dots \mu_{4-k}},\\
    \sigma\cdot\tau = -\ast (\sigma\wedge\ast \tau) =
    \frac{1}{k!} \sigma^{\mu_1\dots\mu_k} \tau_{\mu_1\dots\mu_k}, \\
    \norm{\sigma}^2 = \sigma\cdot\sigma =
    \frac{1}{k!} \sigma^{\mu_1\dots\mu_k} \sigma_{\mu_1\dots\mu_k},
\end{gather}
where, according to the Einstein
summation convention, repeated indices are summed from $0$ to $3$,
and $\epsilon$ is the volume form of $g$. Note that $\tr T_F=0$, hence
taking the trace of Equations~\eqref{eq:einstein} yields
$R_g=0$.  Thus, Equation~\eqref{eq:einstein} can
be rewritten as
\begin{equation}    \label{eq:ricci}
    \Ric_g = 2\, T_F.
\end{equation}
We are using rationalized units in which $4\pi G=1$,
where $G$ is the gravitational constant.
We will seek asymptotically flat, stationary and axially symmetric
solutions of
Equations~\eqref{eq:einstein}--\eqref{eq:maxwell2}.

These equations reduce,
using an idea originally due to Ernst~\cite{ernst},
to an axially symmetric harmonic map $\vp\colon\R^3\bs\Sigma\to\HC$,
where $\Sigma=\A\bs\bigcup_{j=1}^N I_j$ consists of the axis of symmetry
$\A\subset\R^3$ with $N$ closed intervals $I_j$
removed, and $\HC = SU(1,2)/S(U(1)\times U(2))$ is the complex
hyperbolic plane, see~\cite{mazur,carter85}.
Each interval $I_j$ corresponds
in $(M,g)$ to a connected component of the event horizon.  Thus
$N$ will denote, throughout this paper, the number of black holes.
The space $\HC$ has negative sectional curvature,
and finite energy harmonic maps into spaces of negative sectional
curvature are well understood, see~\cite{schoen82,schoen83}.
Nevertheless, this problem is analytically interesting due to the fact
that the boundary conditions for $\vp$ on $\Sigma$ and as $r\to\infty$ are
singular, and hence $\vp$ has infinite energy.
For the case $N=1$, there is a
family of solutions known in closed form, the Kerr-Newman black holes,
see~\cite{carter73}.  We note that the problem
has $4N-1$ parameters:
$N$ masses $m_j$, $N-1$ distances $d_j$, $N$ angular momenta $L_j$,
and $N$ electric charges $q_j$.

Special cases of the Ernst reduction have been used to prove
non-existence results for the Einstein equations.  In early work
along this line, Weyl investigated the vacuum static
case and showed that
the equations reduce to a single linear equation.  The case $N=1$ was
known to have the Schwarzschild solution.
Superimposing two such solutions, Weyl
obtained new solutions which could be interpreted as
equilibrium configurations of a pair of black holes.  However, with
Bach~\cite{bachweyl} in 1921, they showed that
an obstruction arose as a conical singularity along the axis
separating the two black holes.  Having interpreted this singularity as
the gravitational force, they computed its value and verified that the
result was asymptotic to the Newtonian gravitational force in the appropriate
limit.  The reduction was also used by Robinson~\cite{robinson},
following work of Carter~\cite{carter73}, to prove that
within the $N=1$ vacuum case, the Kerr solutions were unique.
Mazur~\cite{mazur},
and Bunting~\cite{carter85} independently,
generalized this uniqueness result to the charged $N=1$ case.

In~\cite{weinstein90,weinstein92}, we used the Ernst reduction to construct
vacuum solutions which could be viewed as
nonlinear generalizations of the Weyl solutions.
Clearly, it is of interest to find out whether
the obstruction found by Weyl in the
vacuum nonrotating case persists also for all possible
choice of the parameters, \ie whether the force between rotating black
holes is always positive.
It has been conjectured for some time that this is so,
see~\cite[Problem 14]{penrose}, but the evidence is limited.
Indeed, it is unlikely that the angle deficiencies can in general
be computed exactly.  We mention here a number of partial results,
\eg~\cite{litian92}, where the small angular
momentum case is treated, \cite{weinstein94}, where
non-existence is proved in the extreme regime, and also~\cite{litian91},
where Y.~Li and G.~Tian proved non-existence in the case when the
solution admits an involutive symmetry.

In the present paper, we begin the generalization of this work to the
Einstein/Maxwell Equations.  We prove the existence of
a unique $(4N-1)$-parameter family of solutions to the reduced problem.
Our main tool is the study of
harmonic maps with prescribed singularities into classical globally
symmetric spaces of noncompact type and rank one~\cite{weinstein95},
\ie the real, complex, and quaternionic hyperbolic spaces,
see~\cite{mostow}.  However, the results of~\cite{weinstein95}
do not apply directly to the problem considered here.
In~\cite{weinstein95}, we restricted our attention to bounded domains
$\Omega\subset\R^n$, and to singular sets $\Sigma$
compactly contained in $\Omega$, while here $\Omega=\R^n$ and
$\Sigma$ is unbounded.  The necessary
generalization is achieved in two steps.  First, we
consider the problem on a large ball $B\subset\R^n$, allowing, however,
the singular set $\Sigma$ to meet $\D B$.  Then, we let $B$ exhaust
$\R^n$.  In this second step, we need the boundary data at infinity
in $\R^n$ to be given by a harmonic map which admits those
singularities which
lie along the unbounded components of $\Sigma$.  In the
application to black holes, this map is obtained from
the Kerr-Newman solutions.

The main motivation for this generalization is an attempt to come to
a better understanding of the force between rotating black holes,
\ie the magnitude of the conical singularities on the bounded components
of the axis.  An important question in this respect is the dependence of
this force on the parameters.  It should be pointed out that
the force can vanish in the case of the Einstein/Maxwell Equations.
Indeed, even in the static case,
equilibrium can be achieved with masses and
charges being equal~\cite{majumdar47,papapetrou47}.
However, these configurations are
extreme, \ie all the event horizons are degenerate,
see Section~\ref{sec:parameters}.  Consequently, one could
still pose the question of
whether the force is positive for all nondegenerate
Einstein/Maxwell black holes~\cite{penrose}.
In a future paper,
we will study the regularity properties of these maps.  This is
a necessary step
in order to complete the application of the results presented here
to black holes.

The plan of the paper is as follows.  In Section~\ref{sec:ernst}, we
review the Ernst Reduction as it applies to the Einstein/Maxwell
Equations.  We carry out the reduction in two steps.  First in
Section~\ref{sec:axial}, we reduce the equations in the presence of one
Killing field.
Then, the second reduction, with two commuting Killing fields, is
carried out in Section~\ref{sec:stat}.  In the rest of
Section~\ref{sec:ernst}, we examine the parameters of the problem,
derive singular boundary conditions, and set-up
a Reduced Problem for the stationary and axially symmetric
Einstein/Maxwell Equations.  In Section~\ref{sec:harmonic}, we prove our
main existence and uniqueness result.  In Section~\ref{sec:einstein}, we
address briefly the problem of going back from a solution of the reduced
problem to a solution of the Einstein/Maxwell Equations.

\section{The Generalized
Ernst Reduction for the Einstein/Maxwell Equations}
\label{sec:ernst}

In this section, we describe the generalized Ernst reduction for the
Einstein/Maxwell equations in the stationary
and axially symmetric case.
The main result in this section
is Theorem~\ref{thm:ernst}, based on which we state the
Reduced Problem for the stationary and axially symmetric Einstein/Maxwell
Equations in terms of harmonic maps with prescribed singularities.
The computations are carried out mostly in the exterior algebra
formalism which is particularly well
suited to the Maxwell Equations.

\subsection{Axial Symmetry} \label{sec:axial}

We first describe the connection between the
axially symmetric Einstein/Maxwell
Equations and harmonic maps into $\HC$.
We note that the arguments given here could be applied to
any spacelike Killing field.

\begin{defn}
Let $(M,g)$ be an
oriented $4$-dimensional Lorentzian manifold, and let $F$ be a
two-form on $M$.
We say that $(M,g,F)$ is \emph{$SO(2)$-symmetric} if
$SO(2)$ acts effectively on $(M,g)$ as a group of
isometries leaving $F$ invariant.  The \emph{axis} in an
$SO(2)$-symmetric spacetime, is the set of points fixed by the action of
$SO(2)$.  We say that $(M,g,F)$ is axially
symmetric if it is $SO(2)$ symmetric, and it has a nonempty axis.
\end{defn}

Let $(M,g,F)$ be a simply connected
$SO(2)$-symmetric solution of the Einstein/Maxwell Equations.
Let $\xi$ be the Killing field generator,
then $\Lie_\xi F=0$.  Note that if $(M,g)$ is
\emph{causal}, \ie admits no closed
causal curves, then $\xi$ either is spacelike, or vanishes.
Define the one-forms $\alpha=i_\xi F$, and $\beta=i_\xi{\ast F}$.
Then, we find
$d\alpha= -i_\xi dF + \Lie_\xi F=0$.  Thus, there is a function $\chi$
such that $d\chi=\alpha$.  Similarly $d\beta=0$, hence there is a function
$\psi$ such that $d\psi=\beta$.
Note that $\chi$ and $\psi$ are determined only up to
constants.  Now, define the one-form
$\gamma=\chi\,d\psi-\psi\,d\chi$, and observe that
$d\gamma=2\alpha\wedge\beta$.

It is easy to see from~\eqref{eq:star} that for any $k$-form $\sigma$
and one-form $\theta$, we have:
\[
    i_\theta {\ast \sigma} = \ast (\sigma\wedge\theta),
\]
where we have used the metric $g$ to identify the tangent space
$T_p M$ and its dual $T_p^*M$.
Using $[\Lie_\xi,\ast ]=0$, and $\ast \ast = (-1)^{k+1}$ on
$k$-forms, it follows that
\[
    \delta(\sigma\wedge\xi) = (-1)^{k+1} \Lie_\xi \sigma +
    \delta\sigma\wedge\xi,
\]
where $\delta={\ast d\ast }$ is the divergence operator on forms.
Thus, if we define the \emph{twist} of $\xi$ by
$\omega={\ast (d\xi \wedge \xi)}$, we find
\[
    \ast d\omega = \delta(d\xi\wedge \xi)
    = - \Lie_\xi d\xi + \delta d \xi\wedge \xi =
    \delta d \xi \wedge \xi.
\]
Now for any Killing field $\xi$, we have $\delta\xi=0$, and
\begin{equation}    \label{eq:killing}
    \delta d \xi = - 2 \tr\nabla^2 \xi = 2 \, i_\xi \Ric_g.
\end{equation}
Hence in view of Equation~\eqref{eq:ricci}, we obtain
$\ast d\omega = 4 \, i_\xi T_F\wedge \xi$.
Furthermore, in view of~\eqref{eq:energy:momentum},
$i_\xi T_F = -i_\alpha F - (1/2) \norm{F}^2 \xi$.
Thus, we have $\ast d\omega = 4 \, \xi\wedge i_\alpha F$.
On the other hand, $\beta = \ast (F\wedge\xi)$, hence
\[
    \ast (\alpha\wedge\beta) = -i_\alpha\ast \beta
    = \xi\wedge i_\alpha F.
\]
It follows that
\begin{equation}    \label{eq:domega}
    d\omega = 2\, d\gamma,
\end{equation}
\ie the one-form $\omega - 2\gamma$ is closed.
We conclude that
there is a function $v$, also determined up to a constant, such that
$2\, dv=\omega - 2\gamma$.

Let $M'=\{x\in M; \> \norm{\xi}^2>0\}$, and let
$u=-\log\norm{\xi}$ on $M'$.  Then, we have $i_\xi d\xi =
-di_\xi\xi + \Lie_\xi \xi = 2 e^{-2u} du$.  Thus, since $\delta du = -
\Delta u$, we have $\delta(e^{-2u}du)=
2e^{-2u} \norm{du}^2 - e^{-2u} \Delta u$.  On the other hand,
$i_\xi d\xi=-\ast (\ast d\xi\wedge\xi)$, hence using
Equations~\eqref{eq:killing}, \eqref{eq:ricci},
and~\eqref{eq:energy:momentum}, we obtain:
\[
    \delta(i_\xi d\xi) = -i_\xi\delta d\xi + \norm{d\xi}^2
    = -2\bigl(\norm{\alpha}^2 + \norm{\beta}^2\bigr) + \norm{d\xi}^2.
\]
Furthermore, since $\omega=i_\xi\ast d\xi$, we find
\[
    \norm{\omega}^2
    = \ast (\ast d\xi\wedge i_\xi d\xi\wedge\xi)
    + \ast (\ast d\xi\wedge d\xi\wedge
        i_\xi\xi)
    = 4e^{-4u} \norm{du}^2 - e^{-2u} \norm{d\xi}^2
\]
\ie $\norm{d\xi}^2 = 4\, e^{-2u} \norm{du}^2 - e^{2u} \norm{\omega}^2$.
We conclude that
\begin{equation}    \label{eq:u}
    \Delta u - \frac12 e^{4u} \norm{\omega}^2 -
    e^{2u}\bigl(\norm{\alpha}^2 + \norm{\beta}^2) = 0.
\end{equation}
Now, $\delta\omega=\ast (d\xi\wedge d\xi)$, and
\begin{align*}
    2\,du\cdot\omega
    & = e^{2u}\ast (d\xi\wedge i_\xi d\xi \wedge\xi) +
        e^{2u} \ast (d\xi\wedge d\xi\wedge i_\xi\xi) \\
    &= -2 \, du\cdot\omega + \ast (d\xi\wedge d\xi),
\end{align*}
\ie $4\, du\cdot\omega = \ast (d\xi\wedge d\xi)$.
Thus, we find that $\delta(e^{4u}\omega) = -4e^{4u} du\cdot\omega +
e^{4u} \delta\omega = 0$, which we write as
\begin{equation}    \label{eq:v}
    \Div(e^{4u} \omega) = 0,
\end{equation}
where we have put
$\Div\sigma = -\delta\sigma$ for any one-form $\sigma$.
In addition, since $\alpha = -\ast (\ast F\wedge\xi)$, we find
$\delta\alpha = -\ast d(\ast F\wedge\xi) = d\xi \cdot F$, and
\begin{align*}
    2\, du\cdot\alpha
    &= -e^{2u}\ast (d\xi\wedge i_\xi\ast F \wedge \xi) -
        e^{2u} \ast (d\xi\wedge \ast F \wedge i_\xi\xi) \\
    &= e^{2u} \beta \cdot\omega + d\xi \cdot F.
\end{align*}
It follows that $\delta(e^{2u}\alpha) = -2e^{2u} du\cdot\alpha +
e^{2u}\delta\alpha = - e^{4u} \beta\cdot\omega$, \ie
\begin{equation}    \label{eq:chi}
    \Div(e^{2u}\alpha) - e^{4u} \beta\cdot\omega = 0.
\end{equation}
Similarly, $\delta(e^{2u}\beta) = e^{4u} \alpha\cdot\omega$, \ie
\begin{equation}    \label{eq:psi}
    \Div(e^{2u}\beta) + e^{4u} \alpha\cdot\omega = 0.
\end{equation}

Substituting the definitions of $v$, $\chi$ and $\psi$ into
Equations~\eqref{eq:u}--\eqref{eq:psi}, it follows that
$\vp=(u,v,\chi,\psi)$ satisfies the following system of equations:
\begin{align}
    \label{eq:hu}
    \Delta u -
    2 e^{4u} \norm{\nabla v
    + \chi \nabla \psi - \psi \nabla \chi}^2
    - e^{2u} \bigl( \norm{\nabla \chi}^2
    + \norm{\nabla \psi}^2 \bigr)
    &= 0 \\
    \label{eq:hv}
    \Div \bigl( e^{4u} (\nabla v + \chi\nabla\psi
    - \psi\nabla\chi) \bigr)
    &= 0 \\
    \label{eq:hc}
    \Div \bigl( e^{2u} \nabla\chi \bigr)
    - 2 e^{4u} \nabla\psi \cdot (\nabla v + \chi\nabla\psi -
    \psi\nabla\chi)
    &= 0 \\
    \label{eq:hp}
    \Div \bigl( e^{2u} \nabla\psi \bigr)
    + 2 e^{4u} \nabla\chi \cdot (\nabla v + \chi\nabla\psi -
    \psi\nabla\chi)
    &= 0.
\end{align}
It is now clear that
for every subset $\Omega\subset\subset M'$,
$\vp=(u,v,\chi,\psi)$ is a critical point of the functional:
\[
    E_\Omega(\vp)=
    \int_{\Omega}\left\{\norm{\nabla u}^2 + e^{4u}\norm{\nabla v +
    \chi\nabla\psi - \psi\nabla\chi}^2 + e^{2u}\bigl(
    \norm{\nabla\chi}^2 + \norm{\nabla\psi}^2\bigr) \right\} d\mu_g,
\]
where $d\mu_g$ is the volume element of the metric $g$.
Thus, if we  choose an `upper half-space model' for
$\HC$, \ie $\R^4$ with the metric given by the line element:
\begin{equation}    \label{eq:hyperbolic}
    ds^2 = du^2 + e^{4u}
    (dv + \chi\,d\psi - \psi\,d\chi)^2
    + e^{2u}
    \bigl( d\chi^2 + d\psi^2 \bigr),
\end{equation}
then the map $\vp\colon(M',g)\to\HC$ is a harmonic map,
see Section~\ref{sec:harmonic} and~\cite{weinstein95}.

\subsection{Stationary and Axially Symmetric Solutions}
\label{sec:stat}

We now turn to the case where $(M,g,F)$ admits an additional
symmetry.

\begin{defn}
Let $(M,g)$ be an oriented and time-oriented
$4$-dimensional Lorentzian
manifold and let $F$ be a two-form
on $M$.  Let $G$ be a group acting on $M$.
The orbit of a point $p\in M$
is \emph{degenerate} if the isotropy subgroup at $p$ is nontrivial.
The \emph{axis} is the set of points whose orbits are degenerate.
We say that $(M,g,F)$ is \emph{stationary and $SO(2)$-symmetric}
if the group $G=\R\times SO(2)$ acts effectively
on $(M,g)$ as a group of isometries leaving $F$
invariant and such that the orbits of points not on the axis
are timelike two-surfaces.
We say that $(M,g,F)$ is \emph{stationary and axially
symmetric} if it is stationary and $SO(2)$-symmetric and has a nonempty
axis.
\end{defn}

Assume that $(M,g,F)$ is stationary and axially symmetric, and
let $\xi$ be the Killing
field generator of the $SO(2)$-symmetry
normalized so that its orbits are
closed circles of length $2\pi\norm{\xi}$.
Let $\tau$ be a linearly independent generator.  Then, we have
$[\xi,\tau]=0$, and as before, if $(M,g)$ is causal, then $\xi$ is either
spacelike or vanishes, \ie $\xi$ is spacelike outside the axis.
We will now prove that if $(M,g,F)$ is a stationary and
axially symmetric solution of the Einstein/Maxwell Equations,
then:
\begin{itemize}
\item[(i)] $F$ and $\ast F$ vanish on the orbits, \ie
       $F(\xi,\tau)=\ast F(\xi,\tau)=0$;
\item[(ii)] the distribution of planes orthogonal to the orbits of
       $G$ is integrable.
\end{itemize}
To see~(i), note that $[\Lie_\tau,i_\xi]=i_{[\tau,\xi]}=0$,
hence we have
\[
     d i_\tau \alpha = -i_\tau d\alpha + \Lie_\tau i_\xi F = 0.
\]
Since $i_\tau\alpha=0$ on the axis it follows that
$F(\xi,\tau)=i_\tau\alpha=0$ everywhere.  Similarly,
$\ast F(\xi,\tau)= i_\tau\beta = 0$.  To show~(ii), it suffices
by Frobenius Theorem to show that:
\begin{equation}    \label{eq:integrable}
    \ast (\xi\wedge\tau \wedge d\xi) = 0, \qquad
    \ast (\xi\wedge\tau\wedge d\tau) =0.
\end{equation}
However, in view of~\eqref{eq:domega} and~(i), we have
\[
    d\ast (\xi\wedge\tau\wedge d\xi) = di_\tau\omega
    = - i_\tau d\omega - \Lie_\tau\omega
    = 4 i_\tau(\alpha\wedge\beta)
    = 0,
\]
where $\omega$ is the twist of $\xi$.
Thus, $\ast (\xi\wedge\tau\wedge d\xi)$ is constant, but since it
vanishes on the axis, it is identically zero.  Similarly,
$\ast (\xi\wedge\tau\wedge d\tau)=0$.

Let $Q$ be an integral surface of the
distribution of planes orthogonal to the orbits, and let $h$
be the metric induced by $g$ on $Q$.
Then, the quotient space $M/G$, with its quotient metric
can be identified with $(Q,h)$.
We choose the orientation on $Q$ so that $\ast(\xi\wedge\tau)$ is positively
oriented.
The map $\vp\colon M'\to\HC$ is invariant
under $G$, hence reduces to a map,
which we also denote by $\vp$, on the quotient
$Q'=Q\cap M'$.  Indeed,
let $\zeta$ be an arbitrary Killing field generator in the Lie algebra of
$G$.  We have
$\zeta e^{-2u} = 2g( [\zeta,\xi] ,\xi)=0$,
and hence $\zeta u =0$.  Also,
$\zeta\chi = i_\zeta \alpha = F(\xi,\zeta)=0$,
and similarly $\zeta\psi=0$.
Finally, $\gamma(\zeta) = \chi \beta(\zeta) -\psi \alpha(\zeta)=0$, hence
$2\, \zeta v= i_\zeta(\omega-2\gamma) = \ast (d\xi\wedge\xi\wedge\zeta)-
2\gamma(\zeta) = 0$.

Define $\sigma=\xi\wedge\tau$, then on $M'$ we have $\norm{\sigma}^2 =
\norm{\xi}^2 \norm{\tau}^2 - (\xi\cdot\tau)^2 < 0$.  Let $\rho^2 =
-\norm{\sigma}^2$, then $\rho$ is invariant under $G$, and thus reduces
to a function on $Q'$.  It follows that for every subset
$\Omega\subset\subset Q'$, the map $\vp\colon
(Q',h)\to\HC$ is a critical point of the functional:
\[
    E'_\Omega(\vp) = \! \int_\Omega
    \left[ \norm{\nabla u}_h^2
    + e^{4u} \norm{\nabla v + \chi\nabla\psi
    - \psi\nabla\chi}_h^2 + e^{2u} \bigl( \norm{\nabla\chi}_h^2 +
    \norm{\nabla\psi}_h^2 \bigr) \right] \rho d\mu_h,
\]
where $\norm{\cdot}_h$ is the norm with respect to the metric $h$,
and $d\mu_h$ is the volume element of the metric $h$.
Indeed, if $\Omega\subset Q'$ is such a subset,
and $f$ is a function on $M'$
invariant under $G$, then $2\pi \int_\Omega f \rho\, d\mu_h =
\int_{G\cdot\Omega}
f\, d\mu_g$, where $G\cdot\Omega$ is the orbit of $\Omega$ under $G$.

Next, we show that $\Delta_h\rho=0$.  Since $\Delta_g \rho = \rho^{-1}
\Div_h (\rho\, \nabla\rho)$, it suffices to prove that $\Delta_g \rho =
\rho^{-1} \norm{d\rho}^2$.  To see this, note first that
$\rho^2 = -\sigma(\xi,\tau) = i_\xi i_\tau \sigma$.
Thus, we obtain:
\begin{equation}    \label{eq:drho}
    2\rho \, d\rho = di_\xi i_\tau\sigma = i_\xi i_\tau d\sigma=
    -\ast  (\ast d\sigma\wedge\sigma).
\end{equation}
Therefore, we see that
\[
    2\, \rho \, \Delta_g \rho + 2\, \norm{d\rho}^2 =
    -\delta(2\, \rho\, d\rho) = \ast  d(\ast d\sigma \wedge \sigma)
    = \sigma \cdot \delta d\sigma - \norm{d\sigma}^2,
\]
or equivalently
\begin{equation}    \label{eq:deltarho}
    \Delta_g\rho = \frac1{2\rho} (\sigma\cdot\delta d\sigma -
    \norm{d\sigma}^2 - 2\norm{d\rho}^2).
\end{equation}
On the other hand, since $d\sigma=d\xi\wedge\tau - \xi\wedge d\tau$, we
have, in view of Equation~\eqref{eq:integrable}, that
$i_\xi\ast  d\sigma=\ast (d\sigma\wedge\xi)=0$, and similarly $i_\tau\ast
d\sigma=0$.  We deduce, using~\eqref{eq:drho}, that
\[
    4 \rho^2 \norm{d\rho}^2
    = \ast ( i_\tau d\sigma \wedge \ast d\sigma \wedge i_\xi\sigma)
    = - \ast (d\sigma \wedge \ast d\sigma \wedge i_\tau i_\xi \sigma)
    = -\rho^2 \norm{d\sigma}^2,
\]
\ie
\begin{equation}    \label{eq:drho2}
    \norm{d\sigma}^2 = - 4\, \norm{d\rho}^2
\end{equation}
Furthermore, we claim that $\sigma\cdot \delta  d\sigma=0$.
Indeed, in view of Equation~\eqref{eq:killing}:
\[
    \delta d \sigma = \ast d (i_\tau \ast d\xi - i_\xi\ast d\tau)
    = \delta d\xi \wedge \tau + \xi \wedge \delta d\tau
    = 2 \, (i_\xi \Ric_g \wedge \tau + \xi \wedge i_\tau \Ric_g),
\]
and consequently,
\[
    \sigma \cdot \delta d\sigma = i_\tau i_\xi \delta d\sigma
    = 2\, \norm{\tau}^2 \Ric_g( \xi,\xi) -
    4\, (\xi\cdot\tau) \Ric_g(\xi,\tau) +
    2\, \norm{\xi}^2 \Ric_g(\tau,\tau).
\]
Introducing $\alphat=i_\tau F$, and $\betat=i_\tau \ast F$, we can
use Equations~\eqref{eq:ricci} and~\eqref{eq:energy:momentum} to write
\begin{align*}
    \Ric_g(\xi,\xi) &= \norm{\alpha}^2 + \norm{\beta}^2 \\
    \Ric_g(\xi,\tau) &= \alpha \cdot \alphat + \beta \cdot \betat \\
    \Ric_g(\tau,\tau) &= \norm{\alphat}^2 + |\betat|^2,
\end{align*}
from which it follows that:
\begin{equation}    \label{eq:sigma}
    \sigma\cdot \delta  d\sigma =
    2\, \norm{\tau}^2 \bigl(\norm{\alpha}^2 + |\beta|^2\bigr)
    - 4\,  (\xi\cdot\tau) \bigl( \alpha\cdot\alphat +
    \beta\cdot\betat \bigr) + 2\, \norm{\xi}^2 \bigl(
    \norm{\alphat}^2 + \norm{\betat}^2 \bigr).
\end{equation}
In addition, we have:
\begin{align*}
    \norm{\xi}^2 F &= \xi\wedge\alpha + i_\xi\ast \beta \\
    \norm{\xi}^2 \ast F &= \xi\wedge\beta - i_\xi\ast \alpha,
\end{align*}
and hence,
\begin{align}
    \label{eq:alphat}
    \norm{\xi}^2 \alphat - (\xi\cdot\tau) \, \alpha
    &= i_\tau i_\xi \ast \beta \\
    \label{eq:betat}
    \norm{\xi}^2 \betat - (\xi\cdot\tau) \, \beta
    &= -i_\tau i_\xi \ast \alpha.
\end{align}
A computation similar to the one leading to~\eqref{eq:drho2} yields:
\[
    \norm{i_\tau i_\xi \ast \alpha}^2 = \rho^2 \norm{\alpha}^2,
    \qquad
    \norm{i_\tau i_\xi \ast \beta}^2 = \rho^2 \norm{\beta}^2.
\]
Thus, taking the norm squared of both sides of~\eqref{eq:alphat}
and~\eqref{eq:betat}, and adding the results, we obtain:
\[
    \norm{\xi}^4 \bigl(\norm{\alphat}^2 + |\betat|^2\bigr)
    - 2\,  \norm{\xi}^2 (\xi\cdot\tau) \bigl( \alpha\cdot\alphat +
    \beta\cdot\betat \bigr) + \norm{\xi}^2 \norm{\tau}^2 \bigl(
    \norm{\alpha}^2 + \norm{\beta}^2 \bigr) = 0,
\]
which, in view of~\eqref{eq:sigma}, implies $\sigma\cdot\delta d\sigma=0$
as claimed.  Substituting this result back into~\eqref{eq:deltarho},
and taking into account~\eqref{eq:drho2}, we
obtain $\Delta_g \rho = \rho^{-1} \norm{d\rho}^2$ as required.

Therefore, if $d\rho\ne0$, the function $\rho$ can be used locally
as a harmonic coordinate for the metric $h$ on
$Q'$.  Let $z$ be a conjugate harmonic coordinate, \ie a function on
$Q'$ such that $\norm{dz}^2=\norm{d\rho}^2$, $dz\cdot d\rho=0$, and
$d\rho\wedge dz$ is positively oriented.  Then,
in the $(\rho,z)$-coordinate system, the metric $h$ takes the form:
\[
    h_{ab} dx^a dx^b = e^{2\lambda} ( d\rho^2 + dz^2 ),
\]
where $\lambda = -\log\norm{d\rho}$.  Let $\h=e^{-2\lambda} h$, then
$\h$ is a flat metric on $Q'$, and since the functional $E'_\Omega$ is
conformally invariant, we have that $\vp$ is a critical point of:
\[
    \tilde E_\Omega(\vp) = \! \int_\Omega
    \left[ \norm{\nabla u}_\h^2
    + e^{4u} \norm{\nabla v + \chi\nabla\psi
    - \psi\nabla\chi}_\h^2
    + e^{2u} \bigl( \norm{\nabla\chi}_\h^2 +
    \norm{\nabla\psi}_\h^2 \bigr) \right] \rho d\mu_\h.
\]
This gives a semilinear elliptic
system of partial differential equations for the four
unknowns $(u,v,\chi,\psi)$.

We have obtained that the metric $g$ must be locally of
the form given by the line element:
\begin{equation}    \label{eq:papapetrou}
    ds^2 = - \rho^2 e^{2u}\, dt^2 +
    e^{-2u} \bigl( d\phi - w\, dt \bigr)^2
    + e^{2\lambda} (d\rho^2 + dz^2),
\end{equation}
where $w=-e^{2u} (\xi\cdot\tau)$.  All the metric coefficients
can be determined from the map $\vp$.   Indeed, $u$ is obtained directly
from $\vp$, and we will now obtain equations for
the gradient of $w$ and $\lambda$.  These
quadratures are to be integrated after
the harmonic map system has been solved.

Define $\eta=\tau+w\xi$, and observe that $\eta\cdot\xi=0$,
while $\norm{\eta}^2= \eta\cdot\tau = - e^{2u} \rho^2$.
Furthermore, we have
$dw = 2\, w\, du + e^{2u} i_\tau d\xi = e^{2u} i_\eta d\xi$.
Since $d\xi = 2\, \xi\wedge du - e^{2u} \ast (\xi\wedge\omega)$,
and $\xi\wedge\eta=\sigma$, we find $i_\eta d\xi
= e^{2u} \ast (\omega\wedge\sigma)$, and hence
\begin{equation}    \label{eq:w}
    i_\tau i_\xi \ast  dw = - \rho^2 e^{4u} \omega
\end{equation}
The operator $\rho^{-1} i_\tau i_\xi \ast$
restricted to forms tangential to $Q$ is
the Hodge
star operator of the metric $h$.  Since in two dimensions, the star
operator restricted to
one-forms is conformally invariant,
$\rho^{-1} i_\tau i_\xi \ast$ is also the Hodge star
operator $\cast$ of the
flat metric $\h$ restricted to one-forms.  Now, on one-forms in two
dimensions $\cast\cast=-1$, hence we can rewrite Equation~\eqref{eq:w} as:
\begin{equation}    \label{eq:dw}
    dw = e^{4u}\rho \cast \omega.
\end{equation}
To derive the equations for $d\lambda$,
we must now use of the components of the Einstein/Maxwell Equations in
the $\rho z$-plane:
\begin{align*}
    \Ric_g(\D_\rho,\D_\rho) - \Ric_g(\D_z,\D_z) &=
    2 \, \bigl( i_{\D_\rho} F \cdot i_{\D_\rho} F -
    i_{\D_z} F \cdot i_{\D_z} F \bigr) \\
    \Ric_g(\D_\rho,\D_z) &= 2 \, i_{\D_\rho} F \cdot i_{\D_z} F.
\end{align*}
A straightforward coordinate computation leads to:
\begin{equation}    \label{eq:dlambda}
    \begin{split}
    \lambda_\rho - u_\rho &=
    \rho\left[ u_\rho^2 - u_z^2 + \frac{1}{4} e^{4u}(\omega_\rho^2 -
    \omega_z^2) + e^{2u} (\chi_\rho^2 - \chi_z^2 + \psi_\rho^2 -
    \psi_z^2) \right] \\
    \lambda_z - u_z &= 2 \rho \left[
    u_\rho u_z + \frac{1}{4} e^{4u} \omega_\rho \, \omega_z +
    e^{2u} (\chi_\rho \chi_z + \psi_\rho \psi_z) \right].
    \end{split}
\end{equation}

\begin{defn}
We collect here a few definitions related to the notion of asymptotic
flatness.
A \emph{data set} $(\Mbar,\gbar,\kbar,\Ebar,\Bbar)$
for the Einstein/Maxwell Equations consists of a $3$-manifold $\Mbar$, a
Riemannian metric $\gbar$ on $\Mbar$, a symmetric twice covariant
tensor $\kbar$ on $\Mbar$, and two vectors
$\Ebar$ and $\Bbar$ on $\Mbar$, satisfying the
\emph{Constraint Equations}:
\begin{alignat*}{2}
    \Rbar - \kbar_{ij}\, \kbar^{ij} + (\tr \kbar)^2 &= \norm{\Ebar}^2
    + \norm{\Bbar}^2, & \qquad
    \nablabar^i\,\kbar_{ij} - \nablabar_j \tr \kbar &=
    2\, \epsilonbar_{ijk}\, \Ebar^j\,\Bbar^k, \\ \qquad
    \nablabar^i\, \Ebar_i &= 0, &
    \nablabar^i\, \Bbar_i &= 0,
\end{alignat*}
where $\epsilonbar$ and $\nablabar$ denote respectively
the volume form and the covariant derivative of $\gbar$.
In addition, we require that $\int_S \gbar(\Bbar,n)=0$ for every closed
$2$-surface $S$ non-homologous to zero in $\Mbar$, where $n$ is the unit
normal to $S$. A $3$-manifold $\Mbar$ is \emph{topologically Euclidean at
infinity}, if there is a compact set $K \subset \Mbar$, such that each
connected component of $\Mbar\bs K$, henceforth called an \emph{end},
is homeomorphic to the complement of a ball in $\R^3$.
The data set $(\Mbar,\gbar,\kbar,\Ebar,\Bbar)$
is \emph{strongly asymptotically flat}
if $\Mbar$ is topologically Euclidean at infinity, and if there is on each
end a coordinate system $(x^1,x^2,x^3)$, in which
$\gbar$, $\kbar$, $\Ebar$, $\Bbar$, take the asymptotic forms:
\begin{alignat*}{2}
    \gbar_{ij} &=
    \left( 1 + \frac{2m}{r} \right) \delta_{ij} + o_2(r^{-3/2}),
    &\qquad
    \kbar_{ij} &= o_1(r^{-5/2}), \\
    \Ebar_i &= \frac{q\, x_i}{r^3} + o_1(r^{-5/2}), &\qquad
    \Bbar_i &= o_1(r^{-5/2}),
\end{alignat*}
where $m$ and $q$ are constants.
Here, a function $f$ is $o_k(r^{-l})$ if $\D^j f = o(r^{-l-j})$
for $j=0,\dots,k$, and $r=\sum (x^i)^2$.  A spacetime is
\emph{globally hyperbolic} if it admits a Cauchy hypersurface, \ie a
hypersurface which intersects every inextendible causal curve exactly
once.  Given a data set $(\Mbar,\gbar,\kbar,\Ebar,\Bbar)$
which is strongly asymptotically flat, the
\emph{Cauchy Problem} for the Einstein/Maxwell Equations,
consists of finding a globally hyperbolic
spacetime $(M,g)$, a two-form $F$, and an embedding
$\iota\colon\Mbar\to M$ such that $(M,g,F)$ is a solution
of~\eqref{eq:einstein}--\eqref{eq:maxwell2},
$\gbar$ and $\kbar$ are respectively the
first and second fundamental forms of $\iota$, and such
that $\Ebar=i_\tau F$, and $\Bbar=i_\tau\ast F$, where $\tau$
is the future pointing unit normal of $\Mbar$ in $M$.
The triple $(M,g,F)$ is called the \emph{Cauchy
development} of the data $(\Mbar,\gbar,\kbar,\Ebar,\Bbar)$.
A triple $(M,g,F)$ will be called \emph{asymptotically flat}
if it is the Cauchy development of strongly asymptotically flat data.
A \emph{domain of outer communications} in
an asymptotically flat spacetime $(M,g)$
is a maximal connected open set $O\subset M$ such that from each point
$p\in O$ there are both future and past directed
timelike curves to asymptotically flat regions of $(M,g)$.
An \emph{event horizon} is the boundary of a domain of
outer communications.
\end{defn}

If $(\widetilde M,g)$ is asymptotically flat and
globally hyperbolic, so is any domain of outer communications $M$
in $(\widetilde M,g)$, since $M$ is then the intersection of future and
past sets.  We will restrict our attention to such a domain.
We may then assert that $(M,g)$ is causal.
We assume that $M$ is simply connected.
If in addition, $(M,g,F)$ is a solution of the Einstein/Maxwell Equations
which is stationary and axially symmetric, then there is a unique
Killing field generator
$\tau$, which is future directed, and such that $\norm{\tau}^2\to-1$
at spacelike infinity.  Defining now $\xi$ and $\rho$ as before,
it can be shown, using $\Delta_h\rho=0$,
that $\rho$ has no critical
points, and hence can be used as a harmonic coordinate on all of $Q'$.
Furthermore, the event horizon $H\subset\widetilde M$ can be
characterized by the conditions: $\norm{\xi}^2>0$ and $\rho=0$.

If $\norm{d\sigma}^2$ vanishes identically on some component $H_0$ of $H$,
then we say that $H_0$ is \emph{degenerate}.
If a component $H_0$ of $H$ is nondegenerate, then
by~\eqref{eq:drho2}, we have $\norm{d\rho}^2=\norm{dz}^2\not\equiv0$
on $H_0$.  We will see in Section~\ref{sec:parameters}
that each component $H_0$ of $H$ is a \emph{Killing horizon} in
the sense that the null generators of $H_0$ are tangent to a Killing vector
$\eta_0$ which is non-zero on $H_0$, see~\cite{carter73,chrusciel94}.
Furthermore, $e^{2u} \norm{d\rho}^2$ tends on $H_0$
to a constant $\kappa_0$, the \emph{surface gravity} of $H_0$,
defined by the equation
$d\bigl(\norm{\eta_0}^2\bigr)=-2\kappa_0\eta_0$ which holds on $H_0$.
Thus, $H_0$ is degenerate if and only if $\kappa_0=0$ and the definition of
a degenerate horizon adopted here
coincides with the standard one, see~\cite{carter73}.
If a component $H_0$ of $H$ is nondegenerate, then
$H_0\cap Q$ is an interval on the boundary of $Q$;
otherwise, it is a point.  Indeed, the intersection of $H_0$ with a
spacelike hypersurface must have the topology of a 2-sphere,
see~\cite{cw94}.
We will assume that the event horizon $H$ consists of $N$
nondegenerate components, which we denote by $H_j$, $1\leq j\leq N$.

Let $\A$ be the $z$-axis in $\R^3$, and consider $Q'\times SO(2)=\R^3\bs\A$
with the flat metric $\rho^2 \, d\phi\otimes d\phi+\h$.
Let $\Sigma=\A\bs\bigcup_{j=1}^N I_j$
where $I_j$ are the open intervals on the boundary of $Q'$
corresponding to the $N$ components $H_j$ of the
event horizon $H$.  Since $\norm{\xi}^2>0$ on $I_j$, the map $\vp$
can be extended continuously across $I_j$.
Let $x\in I_j$, then since $I_j$ is of codimension $2$,
one can show, using a cut-off as in Lemma~\ref{lemma:max}, that
$\vp$ is weakly harmonic and has finite energy in a
sufficiently small ball centered at $x$.
It then follows from regularity theory for harmonic
maps~\cite{schoen82,schoen83} that $\vp$ is smooth in that ball.  Thus,
$\vp\colon\R^3\bs\Sigma\to\HC$ is an axially symmetric harmonic map.
We have proved:
\begin{thm} \label{thm:ernst}
Let $(M,g,F)$ be
a solution of the Einstein/Maxwell Equations which is
stationary and axially symmetric.
Assume that $M$ is simply connected, that $(M,g)$ is the domain of outer
communications of an asymptotically flat globally
hyperbolic spacetime, and that every component of the event horizon in
$(M,g)$ is nondegenerate.
Define $\vp=(u,v,\chi,\psi)$, and $\Sigma$
as above.  Then $\vp\colon\R^3\bs\Sigma\to\HC$ is an axially
symmetric harmonic map.  In addition, the metric $g$ is of the
form~\eqref{eq:papapetrou}, and the metric coefficients $w$ and $\lambda$
satisfy Equations~\eqref{eq:dw} and~\eqref{eq:dlambda}.
\end{thm}

\subsection{Physical Parameters}    \label{sec:parameters}

Let $S$ be an oriented
spacelike two-sphere in $M$.  We define the \emph{mass} $m(S)$,
\emph{angular momentum} $L(S)$, and \emph{charge} $q(S)$,
contained in $S$ by:
\begin{align}
    \label{eq:mass}
    m(S) &= - \frac1{8\pi} \int_S \ast d\tau, \\
    \label{eq:angularmomentum}
    L(S) &= \frac1{16\pi} \int_S \ast d\xi, \\
    \label{eq:chargedef}
    q(S) &= - \frac1{4\pi} \int_S \ast F.
\end{align}
The mass and the angular momentum
are standard Komar integrals associated with the
Killing fields $\xi$ and $\tau$, see~\cite{carter73}, and the definition
of the charge is classical.
We will always perform the integration over a submanifold of revolution,
so that the integral can be computed on the quotient $Q$.
Let $1\leq k\leq 3$, let
$\theta$ be a $k$-form and let $\Lambda$
be a $k$-dimensional submanifold in
$M$ which is the orbit of a $(k-1)$-dimensional submanifold
$\Gamma\subset Q$ under the subgroup $SO(2)\subset G$, \ie
$\Lambda=SO(2)\cdot\Gamma$, then we find:
\begin{equation}    \label{eq:orbit}
    \int_\Lambda \theta = - 2\pi \int_\Gamma i_\xi \theta,
\end{equation}
It is well-known that
Stokes' Theorem can used in~\eqref{eq:mass}--\eqref{eq:chargedef} to
relate the conserved quantities of two spheres $S$ and $S'$ which are
homologous in $M$.

Indeed, suppose that
$S$ and $S'$ satisfy $\D\Lambda=S-S'$ for some
$3$-dimensional submanifold $\Lambda$, then:
\begin{equation}    \label{eq:charge}
    q(S) - q(S') = - \frac1{4\pi} \int_{\Lambda} d\ast F=0,
\end{equation}
so that the charge $q(S)$ actually depends only on the homology class
of $S$.  In our situation, the second
homology group is generated by spacelike cross-sections $S_j$ of the
$N$ components $H_j$ of the event horizon $H$.
Hence $q(S)$ is determined by the homology class of $S$ and by
the $N$ parameters $q_j=q(S_j)$, $j=1,\dots,N$,
which we may call the electric charges
of the $N$ black holes.  By~\eqref{eq:orbit}, we find:
\[
    q_j = \frac12 \int_{I_j} d\psi.
\]
In view of~\eqref{eq:charge},
$\psi$ is constant along each component of $\Sigma$.
Let $\Sigma_j$, $j=1,\dots,N+1$,
denote the $N+1$ connected components of $\Sigma$ in order of
increasing $z$.  We may assume
that $q_0=\psi|_{\Sigma_1}=0$, and $\psi|_{\Sigma_j}=q_{j-1}$ for
$j=2,\dots,N+1$.  According to Equation~\eqref{eq:maxwell1}, we also have:
\[
    \frac12 \int_{I_j} d\chi =
    \frac1{4\pi} \int_{S_j} F = \frac1{4\pi} \int_{\D S_j} A = 0,
\]
which simply expresses the absence of magnetic charge.
As a consequence, $\chi$ is constant throughout $\Sigma$, and we may
as well assume $\chi|_\Sigma=0$.

In analogy,
we define the angular momenta:
\begin{equation}    \label{eq:angular}
    L_j = L(S_j) = \frac1{16\pi} \int_{S_j} \ast d\xi
    = \frac18 \int_{I_j} \omega =
    \frac14 \int_{I_j} (dv + \gamma).
\end{equation}
However, since the electromagnetic field carries angular momentum
$L(S)$ is not determined by these parameters.
Indeed, if $\D\Lambda=S-S'$ and if $\Lambda=SO(2)\cdot\Omega$, then:
\[
    L(S) - L(S') = \frac1{16\pi} \int_\Lambda d\ast d\xi =
    \frac12 \int_{\Omega} \alpha \wedge \beta.
\]
Nevertheless, since
$\gamma$ vanishes on $\Sigma$, it is easy
to see from~\eqref{eq:angular},
that $v$ is constant on each component of $\Sigma$.
We introduce the $N$ parameters
$\lambda_j = \int_{I_j} dv$, and note that $L_j
= (\lambda_j + l_j)/4$ where $l_j = \int_{I_j} \gamma$.  We may assume
that $\lambda_0=v|_{\Sigma_1}=0$, and $v|_{\Sigma_j}=\lambda_{j-1}$ for
$j=2,\dots,N+1$.

Finally, define the $N$ masses:
\[
    m_j = m(S_j) = \frac14 \int_{I_j} i_\xi \ast d\tau.
\]
Equation~\eqref{eq:dw} implies that $w$ is constant on each
$I_j$.  The $N$ constants $w_j=w|_{I_j}$ may be interpreted as the
\emph{angular velocities} of the $N$ black holes.  Now, in view
of~\eqref{eq:drho}, we have:
\[
    dz = \cast d\rho = \frac12 \ast d\sigma,
\]
and therefore,
\begin{equation}    \label{eq:dtau}
    i_\xi \ast d\tau = -2 \, dz + \ast (d\xi\wedge\tau).
\end{equation}
Since $d\xi = 2\,\xi\wedge du - e^{2u} \ast(\xi\wedge\omega)$, we have
\begin{equation}    \label{eq:dxi}
    \ast(d\xi\wedge\tau) = -2\,\rho \cast du - w\omega.
\end{equation}
Combining Equations~\eqref{eq:dtau} and~\eqref{eq:dxi}, we obtain:
\[
    i_\xi \ast d\tau = -2\, dz - w \omega - 2\, \rho \cast du.
\]
The last term vanishes on $I_j$, and thus we obtain:
\[
    m_j =
    \frac14 \int_{I_j} (2\, dz + w\omega) = \mu_j + 2 w_j L_j,
\]
where $2\, \mu_j = \int_{I_j} dz$
is the length of $I_j$ in the metric $\h$.
Again, $m(S)$ is not determined by these,
since the electromagnetic field carries energy.
Let $\D\Lambda=S-S'$ where $\Lambda=SO(2)\cdot\Omega$, then we find:
\begin{align*}
    m(S) - m(S') &= \frac14\int_\Omega d(w\omega + 2\rho\star du) \\
    &= \frac12 \int_\Omega \left\{
    e^{2u} ( \norm{\alpha}^2 + \norm{\beta}^2)
    \rho\, d\rho\, dz + 2 w \alpha\wedge\beta \right\}.
\end{align*}
For future reference, we also introduce the $N-1$ parameters
$r_j=\int_{\Sigma_j} dz$, $j=2,\dots,N$.

We conclude this section by verifying as mentioned earlier that $H_j$ is
degenerate if and only if its surface gravity vanishes.
If we let $\eta_j=\tau+w_j\xi$, then $\eta_j$ is a Killing field
tangent to $H_j$, which is
null but non-zero on $H_j$.  Thus, $H_j$ is a Killing horizon.
Let $\kappa_j$ be the surface gravity of $H_j$, then
\[
    \frac{\norm{d\bigl(\norm{\eta_j}^2\bigr)}^2}{4\norm{\eta_j}^2} \to
    \kappa_j^2,
\]
on $H_j$, see~\cite[p.~150]{carter73}.  However, $\norm{\eta_j}^2 = -\rho^2
e^{2u} + e^{-2u} (w - w_j)^2$, hence in view of~\eqref{eq:dw},
\[
    \norm{d\bigl( \norm{\eta_j}^2 \bigr)}^2
    = 4 \rho^2 e^{4u} \norm{d\rho}^2 + O(\rho^4).
\]
Therefore we conclude that
$e^{2u} \norm{d\rho}^2 \to \kappa_j$ on $H_j$ as claimed in
Section~\ref{sec:axial}, see also~\cite[p.~192]{carter73}.

\subsection{Boundary Conditions}    \label{sec:bound}

In order to
complete the reduction, and formulate a reduced problem for the
stationary and axially symmetric Einstein/Maxwell Equations,
we need to derive boundary conditions for $\vp$
on $\Sigma$ and as $r\to\infty$ in $\R^3$, where $r=\sqrt{\rho^2+z^2}$.
Since $u$ behaves like $\log\rho$ near
$\Sigma$ and as $r\to\infty$, these boundary conditions will be singular.
They may be viewed as prescribed singularities
for the map $\vp$ on $\Sigma$ and at infinity.
In particular, $\tilde E_{\R^3}(\vp)$ the
total energy of $\vp$ will be infinite.

Let $u_0$ be the potential of a uniform charge distribution of $\Sigma$,
normalized so that $u_0+\log\rho\to0$ as $r\to\infty$ in $\R^3$.
The map $(u_0,0,0,0)$ is the solution of the
corresponding problem when
$\lambda_j=q_j=0$, \ie the \emph{Weyl solution}.
Define $N+1$ singular harmonic maps
$\vp_j=(u_0,\lambda_{j-1},0,q_{j-1})\colon\R^3\bs\Sigma\to\HC$,
$j=1,\dots,N+1$.  Note that each maps into a geodesic
$\gamma_j(t)=\bigl( t,\lambda_j,0,q_j)$ of $\HC$.  Furthermore,
the distance $\dist(\vp(x),\vp_j(x))$ remains bounded in a
neighborhood of $\Sigma_j$.   Indeed, if $x\in\Sigma_j$ is not an end
point of $\Sigma_j$, then in a ball $B$ small enough about $x$, the
function $u + \log\rho$ is bounded, and the functions $v$, $\chi$,
and $\psi$ are even functions of $\rho$.  It is straightforward now
using~\eqref{eq:hyperbolic} to check that $\dist(\vp,\vp_j)$ is bounded
on $B$.  A similar argument can be used at the end points.
Finally, in order to get a
uniform estimate on $\Sigma$, one uses the asymptotic flatness
of $(M,g,F)$, see the boundary conditions in~\cite{carter73}.
This motivates the following definition~\cite{weinstein95}.
\begin{defn}
Let $\Hy$ be a real, complex, or quaternionic hyperbolic space, let
$\Omega\subset\R^n$ be a smooth bounded domain,
let $\Sigma\subset\Omega$
be a closed smooth submanifold of codimension at least $2$,
possibly with boundary, let $\Sigma'\subset\Sigma$, and
let $\vp,\vp'\colon\Omega\bs\Sigma\to\Hy$ be harmonic maps.
We say that $\vp$ and $\vp'$ are
\emph{asymptotic near $\Sigma'$} if there is a neighborhood
$\Omega'$ of $\Sigma'$ such that $\dist(\vp,\vp')\in
L^\infty(\Omega'\bs\Sigma')$.
\end{defn}
Thus, in this terminology, $\vp$ is asymptotic to $\vp_j$ near
$\Sigma_j$.

Now let $\Sigmat=\Sigma_1\cup\Sigma_{N+1}$ be the union of the unbounded
components of $\Sigma$.
$\R^3\bs\Sigmat$ is the domain of a harmonic
map $\vpt$ corresponding to a Kerr-Newman solution
with one black hole, such that $\vpt$ is asymptotic to $\vp_1$ near
$\Sigma_1$, and asymptotic to $\vp_{N+1}$ near $\Sigma_{N+1}$.
The map $\vpt$ can be written down explicitly.
Introduce the global parameters $q$, $\mu$, $m$, and $a$ defined by:
\[
    q = \sum_{j=1}^N q_j, \quad
    2 \mu = 2 \sum_{j=1}^N \mu_j + \sum_{j=1}^{N-1} r_j, \quad
    ma = \frac14 \sum_{j=1}^N \lambda_j, \quad
    m^2 - a^2 - q^2 = \mu^2.
\]
Here $q$ is the total charge, $2\mu$ is the length,
in the Euclidean metric of $\R^3$, of the interval of the
$z$-axis from the south pole of $I_1$ to the north pole of $I_{N+1}$,
$m$ and $a$ are the mass and angular momentum per mass of the
corresponding Kerr-Newman spacetime.  Define
the Schwarzschild-type elliptical coordinates $(s,\theta)$
on $\R^3$ defined by:
\[
    \rho^2 = (s-m+\mu)(s-m-\mu)\sin^2\theta, \quad
    z = (s-m) \cos\theta,
\]
where we have assumed without loss of generality
that the point $z=0$ on $\A$
is the center of the interval $\tilde I=\A\bs\Sigmat$.
Then, $\vpt=(\tilde u,\tilde v,\tilde \chi,\tilde \psi)$
is given by:
\begin{equation}    \label{eq:kerr-newman}
    \begin{aligned}
    \tilde u &= \log\sin\theta + \frac12 \log\left[ s^2 + a^2 + a^2
    (2ms-q^2)(s^2+a^2\cos^2\theta)^{-1}\sin^2\theta \right] \\
    \tilde v &= ma\cos\theta(3-\cos^2\theta) -
    \frac{a^2 (q^2 s - ma^2\sin^2\theta) \cos\theta\sin^2\theta}{s^2
    + a^2 \cos^2\theta} + 2ma \\
    \tilde \chi &= -qas (s^2 + a^2\cos^2\theta)^{-1}\sin^2\theta \\
    \tilde \psi &= q (s^2 + a^2\cos^2\theta)^{-1}
    (s^2 + a^2) \cos\theta + q.
    \end{aligned}
\end{equation}
Furthermore, the asymptotic flatness of $(M,g,F)$ implies that
$\dist(\vp,\vpt)\to0$ as $r\to\infty$ in $\R^3$, see~\cite{carter73}.

\begin{defn}
Assume that $\Sigma\subset\R^n$, and let
$\vp,\vp'\colon\R^n\bs\Sigma\to\Hy$ be harmonic maps.
We say that $\vp$ and $\vpt$ are \emph{asymptotic at infinity},
if $\dist(\vp,\vpt)\to0$ as $x\to\infty$ in $\Omega\bs\Sigma$.
\end{defn}

We can now formulate the reduced problem for the
stationary and axially symmetric Einstein/Maxwell Equations.
Let $4N-1$ parameters
\begin{equation}    \label{eq:parameters}
    (\mu_1,\dots,\mu_N, r_1,\dots,r_{N-
    1},\lambda_1,\dots,\lambda_N,q_1,\dots q_N)
\end{equation}
be given, with $\mu_j,r_j>0$.  Construct the
set $\Sigma=\A\bs\bigcup_{j=1}^N I_j$, where $I_j$ are $N$ intervals
on the $z$-axis, of lengths $2\mu_j$, distance $r_j$ apart, and let
$\Sigmat=\Sigma_1\cup\Sigma_{N+1}$.  Define the
harmonic maps $\vp_j=(u_0,\lambda_j,0,q_j)$, and the map
$\vpt\colon\R^3\bs\Sigmat\to\Hy$ according to
Equations~\eqref{eq:kerr-newman}.
We associate with the given parameters~\eqref{eq:parameters},
the data $(\vp_1,\dots,\vp_{N+1},\vpt)$
consisting of these $N+2$ singular harmonic maps.
The $4N-1$ physical parameters, $m_j$, $d_j$, $L_j$, and $q_j$,
are determined only a posteriori.  Note that the
parameters~\eqref{eq:parameters} are nonetheless geometric invariants.

\begin{prob}
Given a set of $4N-1$ parameters as in~\eqref{eq:parameters},
prove the existence of a unique axially symmetric harmonic
map $\vp\colon\R^3\bs\Sigma\to\HC$ such that $\vp$ is asymptotic to
$\vp_j$ near $\Sigma_j$ for each $j=1,\dots,N+1$,
and such that $\vp$ is asymptotic to $\vpt$ at infinity.
\end{prob}

This problem will be solved in Section~\ref{sec:harmonic}, see
the Main Theorem and its Corollary.
Clearly, of at least equal importance is the converse: given
a solution of the Reduced Problem, is there a corresponding solution of
the Einstein/Maxwell Equations which is stationary and axially symmetric,
and which is a simply connected domain of outer communications of a
globally hyperbolic and
asymptotically flat spacetime with $N$ nondegenerate
components to the event horizon?
This question will be examined briefly in Section~\ref{sec:einstein}.

\section{Harmonic Maps with Prescribed Singularities}
\label{sec:harmonic}

In this section, we solve the Reduced Problem for the
stationary and axially symmetric Einstein/Maxwell
Equations.  We state a somewhat more general result, whose proof, however,
requires no additional effort.

Let $\Omega\subset\R^n$, and let
$\Sigma$ be a smooth closed submanifold of $\R^n$ of
codimension at least $2$.
Let $\Hy$ be one of the classical globally symmetric space of noncompact
type and rank one,
\ie either real, complex, or quaternionic hyperbolic space of
real dimension $m$.  For our purpose, we take a \emph{harmonic
map} $\vp=(\vp^1,\dots,\vp^m)\colon\Omega\bs\Sigma\to\Hy$ to mean a map
$\vp\in C^\infty(\overline\Omega\bs\Sigma;\Hy)$, which is
for each $\Omega'\subset\subset\Omega\bs\Sigma$
a critical point of the energy functional:
\[
     E_{\Omega'} = \int_{\Omega'} \norm{d\vp}^2,
\]
where $\norm{d\vp}^2=\sum_{k=1}^n
\inner{\D_k\vp}{\D_k\vp}$ is the energy density, and $\inner{\cdot}{\cdot}$
denotes the metric of $\Hy$.  This is equivalent to requiring $\vp$ to
satisfy in $\Omega\bs\Sigma$ the following elliptic system of
nonlinear partial differential equations:
\[
    \Delta \vp^a +
    \sum_{k=1}^n \Gamma^a_{bc}(\vp)\, \D_k\vp^b \, \D_k\vp^c
    = 0,
\]
where $\Gamma^a_{bc}$ are the Christoffel symbols of $\Hy$.

\begin{defn}
Let $\Omega\subset\R^n$ be a smooth domain, and let
$\Sigma$ be a smooth closed submanifold of
$\Omega$ of codimension at least $2$.
Let $\gamma$ be a geodesic in $\Hy$.
We say that a harmonic map $\vp\colon\Omega\bs\Sigma\to\Hy$ is a \emph{
$\Sigma$-singular map into $\gamma$} if
\begin{itemize}
    \item[(i)] $\vp(\Omega\bs\Sigma)\subset\gamma(\R)$
    \item[(ii)] $\vp(x)\to\gamma(+\infty)$ as $x\to\Sigma$
    \item[(iii)] There is a constant $\delta>0$
    such that $\norm{d\vp(x)}^2 \geq \delta \dist(x,\Sigma)^{-2}$
    in some neighborhood of $\Sigma$.
\end{itemize}
\end{defn}

We will assume that we are given $N+1$ disjoint smooth
closed connected submanifolds $\Sigma_j\subset\R^n$ of codimension
at least $2$, possibly with boundary.  Let $J=\{1\leq j\leq
N+1\colon\> \text{$\Sigma_j$ bounded} \}$, and let $\Sigmat=\bigcup_{j\notin
J} \Sigma_j$ be the union of the unbounded components of $\Sigma$.
For $p\in\Hy$, and $\gamma$ a geodesic in $\Hy$, let
$\dist(p,\gamma)$ denote the distance from $p$ to $\gamma(\R)$, \ie
$\inf_t \dist\bigl(p,\gamma(t)\bigr)$.

\begin{defn}    \label{defn:data}
\emph{Singular Dirichlet data} for the harmonic map problem with
prescribed singularities on $\R^n\bs\Sigma$ into $\Hy$ consists of
$N+2$ harmonic maps $(\vp_1,\dots,\vp_{N+1},\vpt)$ satisfying:
\begin{itemize}
\item[(i)]
$\vp_j\colon\R^n\bs\Sigma\to\Hy$
is $\Sigma$-singular into a geodesic $\gamma_j$ of
$\Hy$.
\item[(ii)]
$\norm{\dist(p_0,\vp_j)-\dist(p_0,\vp_{j'})}\leq C$,
for all $1\leq j,j' \leq N+1$,  where $p_0$ is some fixed point of
$\Hy$.
\item[(iii)]
$\vpt\colon\R^n\bs\Sigmat\to\Hy$
is asymptotic to $\vp_j$ near $\Sigma_j$ for $j\notin J$.
\item[(iv)]
$\dist(\vpt(x),\gamma_j)\to0$ as $x\to\Sigma_j$, for $j\notin J$
outside a compact subset of $\R^n$.
\end{itemize}
\end{defn}

The following result is our main existence theorem.

\begin{mainthm}
Let $(\vp_1,\dots,\vp_{N+1},\vpt)$ be singular Dirichlet data on
$\R^n\bs\Sigma$.  Then there exists a unique
harmonic map $\vp\colon\R^n\bs\Sigma\to\Hy$,
such that for each $j=1,\dots,N+1$, $\vp$ is asymptotic to $\vp_j$
near $\Sigma_j$, and such that $\vp$ is asymptotic to $\vpt$ at infinity.
\end{mainthm}

In other words, given a harmonic map $\vpt$ with the correct
asymptotic behavior at infinity, we can modify it to a harmonic map
$\vp$ with prescribed singular behavior near the compact components
of $\Sigma$.  It is easily seen that the data
$(\vp_1,\dots,\vp_{N+1},\vpt)$
defined in Section~\ref{sec:bound} constitute singular Dirichlet data on
$\R^3\bs\Sigma$.
Thus, as an immediate corollary to the Main Theorem, we obtain:

\begin{cor}
The Reduced Problem for the stationary and axially symmetric
Einstein/Maxwell Equations has a unique solution.
\end{cor}

We note that the axial symmetry of the solution $\vp$
follows immediately from the uniqueness statement in the Main
Theorem.

In~\cite{weinstein92}, the case $\Hy=\HR$ corresponding
to the vacuum equations was studied, and in~\cite{weinstein95} we proved
a version of this theorem for maps $\vp\colon\Omega\bs\Sigma\to\Hy$
where $\Omega$ is bounded and $\Sigma$ is compactly contained in $\Omega$.
The proof of our Main Theorem,
which combines the ideas in both these papers, is divided into three
steps.  In step one we consider the problem on a ball $\Ball\subset\R^n$,
but allow $\Sigma$ to extend to $\D\Ball$.  Besides some minor
technical points there are no new difficulties in
this step.  In step two, we use the known map $\vpt$ to
obtain pointwise a priori bounds on $d(\vp,\vp_j)$ and $d(\vp,\vpt)$
which are independent of the size of
$\Ball$.  This allows us to complete the proof in step three.

\subsection{Preliminaries}

The primary tools for the proof of the Main Theorem
are the \emph{Busemann} functions on Cartan-Hadamard manifolds,
see~\cite{eberlein73,heintze77}.  We note that $\Hy$ is
such a manifold with pinched sectional curvatures $-4
\leq \kappa \leq -1$.
Let $\gamma$ be a geodesic in $\Hy$, then
$f_\gamma(p)$, the Busemann
function associated with $\gamma$ evaluated at $p\in\Hy$, is
the \emph{renormalized} distance between $p$ and the
ideal point $\gamma(+\infty)\in\D\Hy$.  It is defined by:
\[
     f_\gamma(p) = \lim_{t\to\infty}
     \left( \dist\bigl(p,\gamma(t)\bigr) - t \right).
\]
It is not difficult to see that the limit is obtained uniformly for
$p$ in compact subsets of $\Hy$.  It is well-known that $f_\gamma$
is analytic, convex, and has gradient of constant
length one.  The level sets, which we denote by
$S_\gamma(t) = \{ p\in\Hy; \> f_\gamma(p) = t \}$,
are called \emph{horospheres}.  They are diffeomorphic to $\R^{m-1}$.
We will also use the \emph{horoballs} defined by $\B_\gamma(t)=\{
p\in\Hy;\> f_\gamma(p) \leq t\}$, and the geodesic balls $\Bh_R(p) =
\{ q\in\Hy;\> \dist(p,q)\leq R\}$.  Two geodesics $\gamma$ and
$\gamma'$ are said to be \emph{asymptotic} if
$\dist\bigl(\gamma(t),\gamma'(t)\bigr)$ is bounded for $t\geq0$.
This is clearly an equivalence relation.  The boundary of $\Hy$ is
defined to be the set of equivalence classes of geodesics in $\Hy$.
We will denote the equivalence class of $\gamma$ in $\Hy$ by
$\gamma(+\infty)$.
Let $\gamma$ be a geodesic in $\Hy$.
We denote the reverse geodesic $t\mapsto\gamma(-t)$ by $-\gamma$,
and we also write $\gamma(-\infty)=-\gamma(+\infty)$.
We will introduce an analytic
coordinate system in $\Hy$ adapted to $\gamma$.  Let $u=f_{-\gamma}$.
Let $v_0\colon\S_{-\gamma}(0)\to\R^{m-1}$ be an analytic
coordinate system on $\S_{-\gamma}(0)$ centered at $\gamma(0)$.  The
integral curves of $\nabla f_{-\gamma}$ are geodesics parameterized by
arclength.  We `drag' the coordinate system $v_0$ along these integral
curves.  More precisely, let $\Phi_t$ be the
analytic flow generated by this vector field,
then $\Phi_{-t}$ maps $\S_{-\gamma}(t)$ to $\S_{-\gamma}(0)$, hence
$v_t=v_0\compose\Phi_{-t}$ is an analytic coordinate system on
$\S_{-\gamma}(t)$.  Define $v\colon\Hy\to\R^{m-1}$ by
$v|_{\S_{-\gamma}(t)}=v_t$, and let $\Phi=(u,v)\colon\Hy\to\R^m$,
then $\Phi$ is an analytic coordinate system on $\Hy$.
In these coordinates, the metric of $\Hy$ reads:
\[
     ds^2 = du^2 +Q_p(dv),
\]
where for each $p\in\Hy$, $Q_p$ is a positive quadratic form on
$\R^{m-1}$.  We note that the convexity of $f_{-\gamma}$ implies that
for each non-zero $X \in\R^{m-1}$, and each fixed $v\in\R^{m-1}$,
$Q_{\Phi^{-1}(u,v)}(X)$ is a positive increasing function of $u$.

The following lemmas were proved
in~\cite{weinstein95}. They are quoted here without proofs.
Except for Lemma~\ref{lemma:coordinates}, they depend only on the bound
$-4\leq \kappa\leq -1$ for the sectional curvatures of $\Hy$.

\begin{lemma}  \label{lemma:expansion}
For any $(u,v)\in\R^m$ and any $X\in\R^{m-1}$, there holds:
\begin{equation}    \label{eq:expansion}
     2 Q_{\Phi^{-1}(u,v)}(X) \leq \frac{\D}{\D u} \left(
     Q_{\Phi^{-1}(u,v)}(X) \right) \leq 4
     Q_{\Phi^{-1}(u,v)}(X).
\end{equation}
\end{lemma}

\vspace{1ex}

\begin{lemma}   \label{lemma:compact}
Let $\gamma$ be a geodesic in $\Hy$.  Then for any $t_0\in\R$, and any
$T\geq0$, we have
\[
    \B_{\gamma}(-t_0+T) \cap \B_{-\gamma}(t_0+T)
    \subset \Bh_R(\gamma(t_0)),
\]
where $R= T + \log2$.
\end{lemma}

\vspace{1ex}

\begin{lemma}   \label{lemma:constant}
Let $\gamma$ and $\gamma'$ be geodesics in $\Hy$, such that
$\gamma(-\infty)=\gamma'(-\infty)$ and
$\gamma(+\infty)\ne\gamma'(+\infty)$.  Then for some $d\in\R$, we have:
\begin{align*}
    \lim_{t\to\infty} \bigl(f_{-\gamma}
    - f_{\gamma}\bigr) \compose\gamma'(t) &=d \\
    \lim_{t\to-\infty} \bigl( f_{-\gamma} + f _{\gamma} \bigr)
    \compose \gamma'(t) &= 0.
\end{align*}
\end{lemma}

\vspace{1ex}

\begin{lemma}   \label{lemma:monotone}
Let $\gamma$ be a geodesic in $\Hy$.  Then,
for any $t_0\in\R$, and any $T\geq (1/2)\log2$, we have
\[
    \inner{\nabla f_\gamma(p)}{\nabla f_{-\gamma}(p)} > 0, \quad
    \forall p\in \S_\gamma(-t_0 + T) \bs \B_{-\gamma}(t_0 + T)
\]
where $\inner{\cdot}{\cdot}$ denotes the metric on $\Hy$.
\end{lemma}

\vspace{1ex}

\begin{lemma}  \label{lemma:coordinates}
Let $\gamma$ be a geodesic in $\Hy$.  Then there is an analytic
coordinate system $\Phi=(u,v)\colon\Hy\to\R^m$ with $u=f_{-\gamma}$
such that in these coordinates, the metric of $\Hy$ is given by:
\[
     ds^2 = du^2 + Q_p(dv),
\]
and satisfies the following conditions:
\begin{itemize}
\item[(i)] Let $R>0$, $t_0\in\R$,
and let $\gamma'$ be a geodesic in $\Hy$ with
$\gamma'(-\infty)=\gamma(-\infty)$.  Then, there exists a
constant $c\geq1$ such that for all $t\geq t_0$, and all
$p\in \Bh_R\bigl(\gamma'(t)\bigr)$, there holds:
\[
     \frac1c \, Q_{\gamma'(t)}(X) \leq Q_p(X) \leq
     c\, Q_{\gamma'(t)}(X), \qquad \forall X\in\R^{m-1}.
\]
\item[(ii)] For all $t,t'\in\R$,
the set $\S_{-\gamma}(t)\cap\B_{\gamma'}(t')$
is star-shaped in these coordinates with respect to its
`center', the point where $\gamma'$ intersects $\S_{-\gamma}(t)$.
\end{itemize}
\end{lemma}

The following lemma is a slight generalization of Lemma~8
in~\cite{weinstein95}.

\begin{lemma}  \label{lemma:max}
Let $\Ball=\Ball_R\subset\R^n$ be the ball of
radius $R$ centered at the origin, and let $\Sigma$
be a smooth closed submanifold of $\R^n$ of codimension
at least $2$.  Suppose that $u\in C^\infty(\overline{\Ball}\bs\Sigma)$
satisfies $\Delta u \geq 0$, and $0\leq u\leq1$ in
$\Ball\bs\Sigma$.  Then, for any $R'<R$, there holds
\begin{equation}
    \label{eq:inH1}
    \int_{\Ball_{R'}} \norm{\nabla u}^2 \leq C,
\end{equation}
where $C$ is a constant that depends only on
$R$, $R'$, and $n$.  Furthermore, we have:
\begin{equation}
    \label{eq:maxprinc}
    \sup_{\Ball\bs\Sigma} u\leq \sup_{\D\Ball\bs\Sigma} u.
\end{equation}
\end{lemma}
\begin{pf}
If $\chi\in C^{0,1}_0(\Ball\bs\Sigma)$ with $0\leq\chi\leq1$,
then integrating by parts the inequality $\chi^2 u\, \Delta u \geq 0$,
and using $0\leq u\leq 1$, we obtain:
\begin{equation}    \label{eq:nabla:u}
     \int_{\Ball} \chi^2\norm{\nabla u}^2 \leq
     4 \int_{\Ball} \norm{\nabla\chi}^2.
\end{equation}
Define the cut-off function $\chi_\epsilon$ by:
\[
     \chi_\epsilon = \begin{cases}
          2-\log r/\log\epsilon & \text{if
          $\epsilon^2\leq r\leq \epsilon$} \\
          0 & \text{if $r\leq\epsilon^2$} \\
          1 & \text{if $r\geq\epsilon$,}
          \end{cases}
\]
where $r(x)=\dist(x,\Sigma)$,
and let $\chi'\in C^\infty_0(\Ball)$ be another cut-off
satisfying $0\leq\chi'\leq1$ and $\chi'=1$ on $\Ball_{R'}$.
Now, take $\chi=\chi_\epsilon\chi'$ in~\eqref{eq:nabla:u},
and then let $\epsilon\to0$ to get:
\[
    \int_{\Ball} (\chi')^2\norm{\nabla u}^2 \leq
    4 \int_{\Ball} \norm{\nabla\chi'}^2,
\]
see~\cite[Lemma~8]{weinstein95},
The estimate~\eqref{eq:inH1} follows immediately.  Now, the same argument
shows that $\Delta u\geq0$ weakly throughout $\Ball$,
hence, by the maximum principle, we obtain~\eqref{eq:maxprinc} over
$\Ball_{R'}$ instead of $\Ball$.  Finally, let $R'\to R$ to
get~\eqref{eq:maxprinc}.
\end{pf}

\subsection{Step One}

In this step, we restrict our attention to a ball
$\Ball\subset\R^n$ large enough so that (iv) in
Definition~\ref{defn:data} holds near $\D\Ball$.
We prove the existence of a solution
$\vp\colon\Ball\bs\Sigma\to\Hy$ in a manner similar to the proof
of~\cite[Theorem~1]{weinstein95}.  We
will assume all the hypotheses of
the Main Theorem.  Some of the elements in the proof of
Proposition~\ref{prop:bounded} will be used also in the next steps.

\begin{prop}   \label{prop:bounded}
Let $\Ball\subset\R^n$ be a large enough ball.  Then there is a unique
harmonic map $\vp\colon\Ball\bs\Sigma\to\Hy$ such that
$\vp$ is asymptotic to $\vp_j$ near $\Sigma_j$ for each $j=1,\dots,N+1$,
and $\vp=\vpt$ on $\D\Ball\bs\Sigma$.
\end{prop}

\noindent\emph{Proof.}
The proof follows closely that of Proposition 2 in~\cite{weinstein95}.
We begin with the uniqueness.  Let $\vp$ and $\vp'$ be two such maps,
then clearly $\dist(\vp,\vp')\in L^\infty(\Ball\bs\Sigma)$,
and $\vp=\vp'$ on $\D\Ball\bs\Sigma$.  Consider the function
$u=\dist(\vp,\vp')^2$.  We have that $\Delta u \geq0$ on
$\Ball\bs\Sigma$, see~\cite{schoen:yau79}, $u$ is bounded on
$\Ball\bs\Sigma$, and $u|_{\D\Ball\bs\Sigma}=0$.  Thus, by
Lemma~\ref{lemma:max}, it follows that $u=0$ and hence
$\vp=\vp'$ on $\Ball\bs\Sigma$.

For the existence proof, we set-up a variational approach.
Pick a smooth open cover of $\R^n$ which separates the $\Sigma_j$'s,
\ie a collection $\{\Omega'_j\}_{j=1}^{N+1}$ of smooth open sets
such that $\bigcup_j \Omega'_j=\R^n$, $\Sigma_j\subset\Omega'_j$,
and $\Sigma_j\cap\Omega'_{j'}=\emptyset$ for $j\ne j'$.
It can easily be arranged that if $j\in J$
then $\Omega_j\subset\subset\Ball$.  Let $\Omega_j=\Omega'_j\cap\Ball$.
Let $\{\chi_j\}_{j=1}^{N+1}$, be a
partition of unity subordinate to the cover
$\{\Omega_j\}_{j=1}^{N+1}$ of $\Ball$ such that
$\chi_j=1$ near $\Sigma_j$.

We may assume, without loss of generality, that all the geodesics
$\gamma_j$ have the same initial point on $\D\Hy$.
Indeed, if this is not the case, we can write
$\vp_j=\gamma_j\compose U_j$, where $U_j$ is a harmonic function on
$\Ball\bs\Sigma$.  Since $\Hy$ has strictly negative curvature,
there are geodesics $\gamma'_j$ with
a common initial point $p_\infty=\gamma'_j(-\infty)\in\D\Hy$
such that $\gamma_j$ and $\gamma'_j$ are asymptotic.  Let
$\vp'_j=\gamma'_j\compose U_j$, then it is easy to see that $\vp_j$ and
$\vp'_j$ are asymptotic near $\Sigma_j$, so that we may replace $\vp_j$
by $\vp'_j$.  We may also assume that the $U_j$'s are all equal.
Indeed, they differ at most by constants, since,
by~(ii) of Definition~\ref{defn:data},
$U_j-U_{j'}$ is a bounded harmonic function on $\R^n$.  So
we may replace $\vp'_j$ by $\vp''_j=\gamma'_j\compose u_0$, where
$U_j=u_0+c_j$ for some $N+1$ constants $c_j$.  Again $\vp''_j$ is
asymptotic to $\vp_j$ near $\Sigma_j$.  Note that the other conditions
in Definition~\ref{defn:data} are not affected by these changes, and
that the singular Dirichlet data $(\vp''_1,\dots,\vp''_{N+1},\vpt)$ is
equivalent to $(\vp_1,\dots,\vp_{N+1},\vpt)$.  We will assume these
changes have already been carried out.

Let $\Phi=(u,v)$ be the coordinate system on $\Hy$ with
$u=f_{-\gamma_1}$ given in Lemma~\ref{lemma:coordinates}.
We will identify any map $\vp\colon\Ball\bs\Sigma\to\Hy$ with its
parameterization $\Phi\compose\vp=(u,v)$ whenever no confusion can
arise.  We have $\vp_j=(u_0,w_j)$ where $w_j\in\R^{m-1}$ are
constants.  We write $u_0=\sum_{j=1}^{N+1} u_j$ where $u_j$ is a
harmonic function on $\R^n$ which is singular only on $\Sigma_j$.
Without loss of generality, we may assume that for $j\in J$,
$u_j(x)=0$ on $\D\Ball$, and in particular $u_j>0$ in $\Ball$.

For any domain $\Omega$, define the norms:
\begin{align}
     \label{eq:norm:u}
     \Norm{u}_{\Omega} &= \left( \int_{\Omega}
     \left\{ u^2 + \norm{\nabla u}^2 \right\}
     \right)^{1/2}, \\
     \label{eq:norm:v}
     \Norm{v}_{\vp_j;\Omega} &= \left( \int_{\Omega}
     \left\{ \norm{v}^2 + Q_{\vp_j}(\nabla v) \right\}
     \right)^{1/2},
\end{align}
where $Q_{\vp_j}(\nabla v)|_x =
\sum_{k=1}^n Q_{\vp_j(x)}\bigl(\nabla_k v(x)\bigr)$ for
$x\in\Omega\bs\Sigma_j$.  We
will use the Hilbert spaces:
\begin{align*}
    H_1(\Omega) &= \{ u\colon\Omega\to\R ; \>
        \Norm{u}_\Omega < \infty \} \\
    H_1^{\vp_j}(\Omega;\R^{m-1}) &= \{ v\colon\Omega\to\R^{m-1}; \>
    \Norm{v}_{\vp_j;\Omega} < \infty \},
\end{align*}
and the subspaces
$H_{1,0}(\Omega)$ and $H_{1,0}^{\vp_j}(\Omega;\R^{m-1})$,
defined to be the closure in $H_1(\Omega)$ of $C^\infty_0(\Omega)$,
and the closure in $H_1^{\vp_j}(\Omega;\R^{m-1})$ of
$C^\infty_0(\Omega\bs\Sigma_j;\R^{m-1})$, respectively.

We write $\vpt=(\ut,\vt)$, and introduce the function $\vt_0$ obtained by
truncating $\tilde v$ near the bounded components of $\Sigma$ and
setting it to its prescribed values $w_j\in\R^{m-1}$:
\[
     \vt_0 = \left( \prod_{j\in J} (1-\chi_j) \right) \vt +
     \sum_{j\in J} \chi_j w_j.
\]
Note that $\vt_0=w_j$ near $\Sigma_j$ for $j\in J$, hence clearly
$\vt_0-w_j\in H_1^{\vp_j}(\Omega_j;\R^{m-1})$.
We also decompose $\uo = \Uo+\Uto$ into
$\Uo = \sum_{j\in J} u_j$\label{page:uo}, the contribution from the
bounded components of $\Sigma$,
and $\Uto = \sum_{j\not\in J} u_j$, the contribution
from the unbounded components.
Clearly, $\Delta \Uo = \Delta \Uto=0$ on
$\Ball\bs\Sigma$ and $\Uo=0$ on $\D\Ball$.
Since $\vpt$ is a harmonic map,
$\ut$ is subharmonic on $\R^n\bs\Sigmat$,
hence $\Delta(\ut-\Uto)\geq0$ on $\R^n\bs\Sigmat$.
Furthermore, $|\ut-\Uto|$ is bounded on the ball
$\Ball_{R+1}$.  Indeed, on the one hand, for $j\not\in J$, we have
$|\ut-\Uto|\leq \norm{\ut-u_0} + \Uo \leq \dist(\vpt,\vp_j) + C_j$
on $\Omega'_j\cap\Ball_{R+1}$ for some constant $C_j$.
On the other hand, for $j\in J$, we have that
both $\ut$ and $\Uto$ are bounded on $\Omega_j$.  Hence it follows from
Lemma~\ref{lemma:max} that $\ut-\Uto\in H_1(\Ball)$.
In addition, $\ut$ satisfies the equation $\Delta\ut =
(\D/\D u)\bigl(Q_{\vpt}(\nabla\vt)\bigr)$, and
Lemma~\ref{lemma:expansion} implies that:
\[
     Q_{\vpt}(\nabla\vt) \leq \frac12 \Delta (\ut-\Uto),
\]
in $\Ball_{R+1}\bs\Sigmat$.  Multiplying by a cut-off $\chi$ as in the
proof of Lemma~\ref{lemma:max} and integrating over $\Ball$,
we conclude, after taking the appropriate limit, that
\begin{equation}    \label{eq:vpt}
     \int_\Ball Q_{\vpt}(\nabla\vt) < \infty.
\end{equation}

Define $\H$ to be the space
of maps $\vp=(u,v)\colon\Ball\bs\Sigma\to\Hy$ satisfying:
\[
     \begin{cases}
     u-\ut-\Uo \in H_{1,0}(\Ball); \\
     v-\vt_0 \in \bigcap_{j=1}^{N+1} H_{1,0}^{\vp_j}(\Ball;\R^{m-1}); \\
     \dist(\vp,\vp_j)\in L^\infty(\Omega_j\bs\Sigma_j), \quad
     \forall j=1,\dots,N+1.
     \end{cases}
\]
For maps $\vp\in\H$ define:
\[
     F(\vp) = \int_\Ball \left\{ \norm{\nabla(u-\uo)}^2 +
     Q_{\vp}(\nabla v) \right\}.
\]
For $R>0$, define $\Hr$ to be the space of maps $\vp\in\H$ such that
$\dist(\vp,\vp_j)\leq R$ for a.e.\
$x\in\Omega_j\bs\Sigma_j$ for $j=1,\dots,N+1$.
We will need the following lemma
which is a direct consequence of Lemma~\ref{lemma:coordinates}.

\begin{lemma}  \label{lemma:equivalent}
Let $R>0$, then there is a $c\geq1$ such that for all $\vp\in\Hr$,
there holds for each $j=0,\dots,N$ and for all $X\in\R^{m-1}$:
\[
     \frac1c\, Q_{\vp_j(x)}(X) \leq Q_{\vp(x)}(X) \leq c\,
     Q_{\vp_j(x)}(X), \quad \text{for a.e.\
     $x\in\Omega_j\bs\Sigma_j$.}
\]
\end{lemma}

We already know that $\int_{\Omega_j} Q_{\vp_j}(\nabla \vt_0) < \infty$ for
$j\in J$, and it follows from Lemma~\ref{lemma:equivalent}
and~\eqref{eq:vpt} that the same holds also for $j\not\in J$.
Indeed, since $\vpt$ is asymptotic to $\vp_j$ near $\Sigma_j$
for $j\notin J$,  there is a constant $c\geq1$ such that
\[
     \int_{\Omega_j} Q_{\vp_j}(\nabla \vt) \leq c \int_{\Omega_j}
     Q_{\vpt}(\nabla \vt) <\infty.
\]
Since $\chi_j$ is smooth and $\nabla\chi_j=0$ near $\Sigmat$, we obtain
that $\int_{\Omega_j} Q_{\vp_j}(\nabla \vt_0) < \infty$ as claimed.
Now, let $\vp=(u,v)\in\H$, then there is
a constant $c\geq1$ such that:
\begin{equation}    \label{eq:Qpart}
     \int_\Ball Q_{\vp}(\nabla v) \leq c \sum_{j=0}^N
     \int_{\Omega_j} Q_{\vp_j}(\nabla v) < \infty.
\end{equation}
Since also $u-\uo = (u-\ut-\Uo) + (\ut-\Uto)
\in H_1(\Ball)$, we conclude that $F$ is
finite on $\H$.

It is straightforward to check that a minimizer $\vp\in\H$ of
$F$ is a harmonic map on $\Ball\bs\Sigma$,
see~\cite{weinstein95}.  Hence, by the regularity theory for
harmonic maps~\cite{schoen82,schoen83}, we have
$\vp\in C^\infty(\overline\Ball\bs\Sigma;\Hy)$.
Furthermore, $\vp$ is asymptotic
to $\vp_j$ near $\Sigma_j$ for $j=1,\dots,N+1$ by construction,
and $\vp=\vpt$ on $\D\Ball\bs\Sigma$.
Thus, to prove Proposition~\ref{prop:bounded}, it
suffices to show that $F$ admits a minimizer in $\H$.

It can be shown, exactly as in~\cite[Proposition~1]{weinstein95},
that $F$ admits a minimizer in $\Hr$.
The main steps are as follows.  Consider a minimizing sequence
$\vph_k=(\uh_k,\vh_k)$ in
$\Hr$.  Then for some subsequence $k'$,
$\uh_{k'}-\uo$ converges weakly in $H_1(\Ball)$
and pointwise a.e.\ in $\Ball$.  Furthermore, using the bound
$\dist(\vph_k,\vp_j)\leq R$ in $\Omega_j$, it is shown that,
perhaps along a further
subsequence, $\vh_{k'}-\vt_0$ also converges weakly
in each $H_{1,0}^{\vp_j}(\Ball,\R^{m-1})$, and pointwise a.e.\ in
$\Ball$.  Let $\vp=(u,v)\in\Hr$ be the weak limit.
Then clearly we have
\begin{equation}    \label{eq:ulsc}
    \int_\Ball\norm{\nabla(u-\uo)}^2 \leq \liminf
    \int_\Ball \norm{\nabla(\uh_{k'} - \uo)}^2.
\end{equation}
Furthermore, $Q_\vp$ is equivalent to $Q_{\vp_j}$ on $\Omega_j$, hence
we have
\begin{equation}    \label{eq:qlsc}
    \begin{split}
    \int_\Ball Q_{\vp}(\nabla v) &=
    \lim \int_\Ball Q_{\vp} (\nabla v,
    \nabla \vh_{k'}) \\
    &\leq \liminf \Biggl[ \left( \int_\Ball \chi_{k'}
    Q_{\vp}(\nabla v) \right)^{1/2} \left( \int_\Ball
    Q_{\vph_{k'}}(\nabla\vh_{k'}) \right)^{1/2} \Biggr],
    \end{split}
\end{equation}
where
\[
    \chi_k = \begin{cases}
        Q_{\vp}(\nabla \vh_k)/Q_{\vph_k}(\nabla \vh_k)
        & \text{if $\vh_k\ne0$} \\
        1 & \text{otherwise}.
        \end{cases}
\]
However $\chi_{k'}\to1$ pointwise a.e.\ in $\Ball$, and $\chi_{k'}$ is
bounded.  Thus, we conclude that
\[
    \lim \int_\Ball \chi_{k'} Q_{\vp} (\nabla v) = \int_\Ball
    Q_{\vp}(\nabla v),
\]
and therefore~\eqref{eq:qlsc} implies:
\[
    \int_\Ball Q_{\vp}(\nabla v) \leq
    \left( \int_\Ball Q_{\vp}(\nabla v) \right)^{1/2}
    \liminf \left( \int_\Ball Q_{\vph_{k'}}(\nabla\vh_{k'})
    \right)^{1/2}.
\]
Combining this with~\eqref{eq:ulsc} we obtain
\[
    F(\vp) \leq \liminf F(\vph_{k'}) = \inf_{\Hr} F,
\]
and it follows that $\vp$ is a minimizer in $\Hr$.

Thus,
it remains to prove that for some $R>0$ large enough,
$\inf_{\H}F=\inf_{\Hr}F$.  This is the content of the next lemma.

\begin{lemma}  \label{lemma:apriori}
There is a constant $R>0$ such that for every $\epsilon>0$, and every
$\vp\in\H$, there is $\vp'\in\Hr$ such that
$F(\vp')\leq F(\vp)+\epsilon$.
\end{lemma}
\begin{pf*}{Proof of Lemma~\ref{lemma:apriori}}
Let $\H^*$ be the space of maps $\vp=(u,v)\in\H$ such that $v=w_j$ in a
neighborhood of $\Sigma_j$ for $j\in J$, $\vp=\vpt$ outside some compact set
$K\subset\Ball$, and $v=w_j$ in a neighborhood of
$\Sigma_j\cap K$ for $j\not\in J$.
Lemma~\ref{lemma:apriori} will follow immediately from: (i) for each
$\vp\in\H$ and each $\epsilon>0$
there is $\vp'\in\H^*$ such that $F(\vp')\leq F(\vp)+\epsilon$;
and (ii) for each $\vp\in\H^*$ there is $\vp'\in\Hr$ such that $F(\vp')\leq
F(\vp)$.  The proof of (i), a standard approximation argument, is
practically unchanged
from~\cite[Lemma~11]{weinstein95}.  We turn to the
proof of (ii).  The bound
$\dist(\vp,\vp_j)\leq R$ will be achieved consecutively
on each $\Omega_j$ beginning with all $j\not\in J$.
The case $j\in J$ is simpler.
Assume without loss of generality that $1\not\in J$.
Introduce a `dual' coordinate system $\bar\Phi=(\ub,\vb)$
with $\ub=f_{\gamma_1}$ as given by Lemma~\ref{lemma:coordinates}.
Write
\[
    ds^2 = d\ub^2 + \Qb_p(d\vb).
\]
for the metric of $\Hy$ in these coordinates.
Set $\ubh = -u_1 + \sum_{j=2}^{N+1} u_j$.  Let $\vp\in\H^*$ and write
$\bar\Phi\compose\vp=(\ub,\vb)$.  Then, since $\norm{\nabla u}^2 +
Q_{\vp}(\nabla v) = \norm{\nabla\ub}^2 + \Qb_{\vp}(\nabla\vb)$,
we find that in $\Ball\bs\Sigma$:
\begin{multline*}
    \norm{\nabla(u-\uo)}^2 + Q_{\vp}(\nabla v) =
    \norm{\nabla(\ub-\ubh)}^2 +\Qb_{\vp}(\nabla\vb) \\
    - 2 \, \nabla(u+\ub) \cdot \nabla u_1
    - 2 \, \sum_{j=2}^{N+1} \nabla(u-\ub) \cdot \nabla u_j
    + 4 \, \sum_{j=2}^{N+1} \nabla u_j \cdot \nabla u_1.
\end{multline*}
The idea is now to integrate this identity over $\Ball$, use the right
hand side to truncate $\ub-\ubh$, and the left hand side to truncate
$u-\uo$.  The convexity of the Busemann functions $u$ and $\ub$ on $\Hy$,
expressed in the monotonicity of $Q$ and $\Qb$, ensures that these
truncations do not increase $F$.  This can be viewed as a weak form of a
variational maximum principle.

Some care, however, needs to be taken because of the
singularities on $\Sigma$.  Integrating first over
$\Ball\bs\Sigma^\epsilon$, where $\Sigma^\epsilon=
\{x\in\Ball;\>\dist(x,\Sigma)\leq\epsilon\}$, we obtain:
\[
    \int_{\Ball\bs\Sigma^\epsilon} \left\{
    \norm{\nabla(u-\uo)}^2 + Q_{\vp}(\nabla v) \right\}
    =\int_{\Ball\bs\Sigma^\epsilon} \left\{
    \norm{\nabla(\ub-\ubh)}^2 +\Qb_{\vp}(\nabla\vb) \right\}
    + I_\epsilon(\vp),
\]
where:
\begin{multline}    \label{eq:Ie}
    \qquad
    I_\epsilon(\vp) = \int_{\Ball\bs\Sigma^\epsilon} \left\{
    - 2  \Div\bigl(
    (u+\ub) \, \nabla u_1 \bigr)  \right. \\
    {} - \left.
    2 \sum_{j=2}^{N+1} \Div\bigl( (u-\ub) \, \nabla u_j \bigr)
    + 4 \sum_{j=2}^{N+1} \nabla u_j \cdot \nabla u_1 \right\}.
    \qquad
\end{multline}
A tedious but straightforward verification shows
that $I_\epsilon(\vp)\to C$ as $\epsilon\to0$, where $C$ is
a constant independent of $\vp\in\H^*$, yielding an integral identity:
\begin{equation}    \label{eq:identity}
    F(\vp)
    =\int_{\Ball} \left\{
    \norm{\nabla(\ub-\ubh)}^2 +\Qb_{\vp}(\nabla\vb) \right\} + C.
\end{equation}
Note that if $\epsilon>0$ is
small enough, we can decompose $\D(\Ball\bs\Sigma^\epsilon)$ into a
disjoint union $\bigcup_{j=1}^{N+1} \D(\Sigma_j^\epsilon)
\cup(\D\Ball\bs\Sigma^\epsilon)$.
Then, for example,
the first term in~\eqref{eq:Ie} can be integrated by parts:
\begin{equation}    \label{eq:Ie1}
    -2 \int_{\Ball\bs\Sigma^\epsilon}
    \Div\bigl( (u+\ub) \, \nabla u_1 \bigr) =
    -2 \left( \int_{\D\Ball\bs\Sigma^\epsilon}
    + \int_{\D\Sigma_1^\epsilon}
    + \sum_{j=2}^{N+1} \int_{\D\Sigma_j^\epsilon} \right)
    (u+\ub) \, \frac{\D u_1}{\D n}.
\end{equation}
For the first integral on the right hand side of~\eqref{eq:Ie1}, we
have, as $\epsilon\to0$:
\begin{equation}    \label{eq:DOmega}
    \int_{\D\Ball\bs\Sigma^\epsilon} (u+\ub) \, \frac{\D u_1}{\D n} \to
    \int_{\D\Ball} (u+\ub) \, \frac{\D u_1}{\D n},
\end{equation}
which clearly is independent of $\vp\in\H^*$.
The integral on the right hand side of~\eqref{eq:DOmega} is finite since
$u+\ub$ is bounded in $\Omega_1\bs\Sigma_1$.  Next,
since $\vp\in\H^*$, there is a compact set $K\subset\Ball$ such that
$\vp=\vpt$ outside $K$ and $\vt=w_1$ in a neighborhood of $\Sigma_1\cap
K$.  We find:
\[
    \int_{\D\Sigma_1^\epsilon} (u+\ub) \, \frac{\D u_1}{\D n} =
    \int_{\D\Sigma_1^\epsilon\bs K} (u+\ub) \, \frac{\D u_1}{\D n},
\]
since for $x\in\D\Sigma_1^\epsilon\cap K$ and $\epsilon>0$ small enough,
$\vp(x)$ lies along the geodesic $\gamma_1$ where $u+\ub=0$.  However for
$x\in \D\Sigma_1^\epsilon\bs K$, we have $\vp(x)=\vpt(x)$ which implies
$\bigl(u(x)+\ub(x)\bigr)\leq\dist(\vpt(x),\gamma_1)\to0$ as
$\epsilon\to0$, by~(iv) of Definition~\ref{defn:data}.
Thus, we obtain
\[
    \int_{\D\Sigma_1^\epsilon\bs K} (u+\ub)
    \, \frac{\D u_1}{\D n} \to 0.
\]
Finally, in view of Lemma~\ref{lemma:constant}, we have that,
for $2\leq j\leq N+1$, $\lim_{\epsilon\to0}(u+\ub)|_{\D_j^\epsilon}=d_j$,
for some constants $d_j$.
Hence the last sum in~\eqref{eq:Ie1} also tends to zero since $u_1$ is
regular in $\Omega_j$ for $j=2,\dots,N+1$.  The other terms
in~\eqref{eq:Ie} are handled similarly.
See~\cite[Proof of Lemma~13]{weinstein95} for more details.

Now, truncate $\ub-\ubh$ above at
\[
    \Tb_1=\sup_{\D\Ball}(\ub-\ubh) + 1,
\]
\ie define a map $\vp'=(u',v')\in\H$
by $\bar\Phi\compose\vp'=(\ub',\vb)$ where
\[
    \ub'-\ubh =\min\{\ub-\ubh,\, \Tb_1\}.
\]
Clearly the map $\vp'\in\H$ and satisfies:
\begin{equation}    \label{eq:inhoroball}
    \vp'(x) \in \B_{\gamma_1}(\ubh(x) + \Tb_1), \qquad
    \forall x\in\Ball\bs\Sigma,
\end{equation}
and also, in view of~\eqref{eq:identity},
$F(\vp')\leq F(\vp)$.
Let $c_1=\sup_{\Omega_1} \sum_{j=2}^{N+1} u_j$, then we obtain
from~\eqref{eq:inhoroball}:
\begin{equation}    \label{eq:inhoroball1}
    \vp'(x)\in \B_{\gamma_1}(-u_1(x) + c_1 + \Tb_1), \qquad
    \forall x\in\Omega_1\bs\Sigma_1.
\end{equation}
Next truncate $u'- \uo$ above at
\[
    T_1 = \max \left\{ \sup_{\D\Ball}(u-\uo) + 1, \,
    \Tb_1, \, (1/2)\log 2 \right\},
\]
to get a map $\vp''=(u'',v')\in\H$ satisfying:
\[
    \vp''(x)\in \B_{-\gamma_1}(\uo(x) + T_1), \qquad
    \forall x\in\Ball\bs\Sigma
\]
and $F(\vp'')\leq F(\vp')$.  As before it follows that:
\begin{equation}    \label{eq:inhoroball2}
    \vp''(x)\in \B_{-\gamma_1}(u_1(x) + c_1 + T_1), \qquad
    \forall x\in\Omega_1\bs\Sigma_1.
\end{equation}
From Lemma~\ref{lemma:monotone} we obtain,
as in~\cite[Lemma~11]{weinstein95}, that the
bound~\eqref{eq:inhoroball1} still holds for $\vp''$:
\begin{equation} \label{eq:inhoroball3}
    \vp''(x)\in \B_{\gamma_1}(-u_1(x) + c_1 + T_1), \qquad
    \forall x\in\Omega_1\bs\Sigma_1.
\end{equation}
Hence, combining~\eqref{eq:inhoroball2} and~\eqref{eq:inhoroball3} with
Lemma~\ref{lemma:compact}, we conclude:
\begin{equation}    \label{eq:bound}
    \vp''(x) \in \Bh_{R_1}\bigl(\vp_1(x)\bigr), \qquad
    \forall x\in\Omega_1\bs\Sigma_1,
\end{equation}
where $R_1=c_1 + T_1 + \log 2$.

Consider now the map $\vp''|_{\cup_{j=2}^{N+1}\Omega_j}$,
and notice that from~\eqref{eq:bound} we can obtain
a pointwise estimate for $\vp''$ on
$\D\bigl( \bigcup_{j=2}^{N+1}\Omega_j \bigr)$.
Thus we can proceed by induction to obtain a map $\vp'''\in\H$ satisfying
$F(\vp''')\leq F(\vp)$, and for each $j=1,\dots,N+1$:
\[
    \vp'''(x) \in \Bh_{R_j} \bigl(\vp_j(x)\bigr), \qquad
    \forall x\in\Omega_j\bs\Sigma_j,
\]
for some constants $R_j$ depending only on the data
$(\vp_1,\dots,\vp_{N+1},\vpt)$,  Setting $R=\max_j R_j$,
we have obtained $\vp'''\in\H_R$ with $F(\vp''')\leq
F(\vp)$.  This completes the proof of Lemma~\ref{lemma:apriori} and of
Proposition~\ref{prop:bounded}.
\end{pf*}

\noindent \emph{Remark.}  It is important to note
that the a priori bounds
$\dist(\vp,\vp_j)\leq R$ in $\Omega_j\bs\Sigma_j$ given by
Lemma~\ref{lemma:apriori} depend on the radius of $\Ball$, hence
cannot be used to obtain a solution on
$\R^n\bs\Sigma$.  This is remedied in the next
step where we obtain a priori bounds independent of the size of $\Ball$.

\subsection{Step Two}   \label{sec:uniform}

In this step, we furnish the main ingredient in
the proof of the Main Theorem.
We establish uniform pointwise a priori bounds for $\dist(\vp,\vpt)$
near $\infty$, and for $\dist(\vp,\vp_j)$ near $\Sigma_j$,
where $\vp\colon\Ball\bs\Sigma\to\Hy$  is the solution
given by Proposition~\ref{prop:bounded}.
As in Proposition~\ref{prop:bounded},
$\Ball\subset\R^n$ is a sufficiently large ball, but
we now use a slightly different open cover of $\R^n$.
For $j\not\in J$, $\vpt$ is
asymptotic to $\vp_j$ near $\Sigma_j$, hence we can choose
$\Omega'_j$ to be a neighborhood of $\Sigma_j$ such that
$\dist(\vpt,\vp_j)$ is bounded on
$\Omega'_j\bs\Sigma_j$.  However, to get an open cover of $\R^n$,
we add another open set $\Omegat'$ which we may choose so that
$\Sigma_j\cap\Omegat'=\emptyset$ for all $1\leq j\leq N+1$.
We set $\Omega_j=\Omega'_j\cap\Ball$,
and $\Omegat=\Omegat'\cap\Ball$.  We also change the normalization of the
harmonic functions $u_j$ for $j\in J$, so that $u_j(x)\to0$ as
$r=\norm{x}\to\infty$ in $\R^n$.

\begin{prop}    \label{prop:uniform}
There is a constant $R>0$ independent of $\Ball$ such that
if $\vp$ is the solution given by Proposition~\ref{prop:bounded}, then
\begin{align}
    \label{eq:atinfinity}
    \dist(\vp,\vpt) &\leq R, \quad\text{in $\Omegat$}, \\
    \label{eq:atj}
    \dist(\vp,\vp_j) &\leq R, \quad\text{in $\Omega_j\bs\Sigma_j$, for
    $j=1,\dots, N+1$}.
\end{align}
\end{prop}
\begin{pf}
Define on $\Ball\bs\Sigma$ the function:
\[
    \nu = \sqrt{1 + \dist(\vp,\vpt)^2} - \Uo,
\]
where recall that $U_0=\sum_{j\in J} u_j$, defined on
page~\pageref{page:uo}, is the contribution to $\uo$ from the bounded
components of $\Sigma$.  Note that since the function
$\dist(\cdot,\cdot)^2$ is convex on $\Hy\times\Hy$, we have
$\Delta\nu\geq0$ on $\Ball\bs\Sigma$, see~\cite{schoen:yau79}.
We claim that $\nu$ is bounded, and
therefore Lemma~\ref{lemma:max} applies.  To see this, fix first
$j\not\in J$, and note that $\vp$ is asymptotic to $\vp_j$,
which is asymptotic to $\vpt$ near $\Sigma_j$.  Hence $\dist(\vp,\vpt)$ is
bounded in $\Omega_j$.  Also, $\Uo$ is bounded in
$\Omega_j$, and hence $\nu$
is bounded there.  Similarly, $\nu$ is bounded on $\Omegat$.
Now, let $j\in J$, and observe
that $\dist\bigl(\gamma_j(0),\vpt\bigr)$ is bounded on $\Omega_j$.  Thus,
since $\dist\bigl(\vp_j,\gamma_j(0)\bigr)=u_j$, we see that
\begin{align*}
    \nu &\leq 1 + \dist(\vp,\vp_j) + \dist\bigl(\vp_j,\gamma_j(0)\bigr)
    + \dist\bigl(\gamma_j(0),\vpt\bigr) - \Uo \\
    &= 1 + \dist(\vp,\vp_j) + \dist\bigl(\gamma_j(0),\vpt\bigr)
    - \textstyle\sum_{\substack{j'\in J\\ j'\ne j}} u_{j'}
\end{align*}
Since $\vp$ is asymptotic to $\vp_j$ near $\Sigma_j$, and $u_{j'}$ is
bounded in $\Omega_j\bs\Sigma_j$ for $j'\ne j$, we obtain $\nu\leq C_j$ in
$\Omega_j\bs\Sigma_j$.  Hence, we conclude
$\nu\leq C$ in $\Ball\bs\Sigma$.  Similarly, we obtain
\[
    \nu \geq -\dist\bigl(\gamma_j(0),\vpt\bigr) - \dist(\vp,\vp_j) -
    \textstyle\sum_{\substack{j'\in J\\ j'\ne j}} u_{j'} \geq -C.
\]
Applying Lemma~\ref{lemma:max},
and using the fact that $\vp=\vpt$ and $U_0=0$ on $\D\Ball\bs\Sigma$,
we deduce that:
\begin{equation}    \label{eq:nu}
    \nu \leq \sup_{\D\Ball\bs\Sigma} \nu \leq 1
\end{equation}
It follows that
\begin{equation}    \label{eq:nubound}
    \dist(\vp,\vpt) \leq \sqrt{\bigl(1 + \Uo\bigr)^2 - 1}.
\end{equation}
Now there is a $T$ such that $U_0\leq T$ on $\bigcup_{j\not\in J}
\Omega'_j \cup \Omegat'$.
Thus, we obtain immediately $\dist(\vp,\vpt)\leq\tilde R$ on $\Omegat$
with $\tilde R = 1+T$.  Furthermore, for $j\notin J$
there are constants $T_j>0$ such that $\dist(\vpt,\vp_j)\leq T_j$ on
$\Omega'_j\bs\Sigma_j$.  Thus, from~\eqref{eq:nubound}, we obtain
the pointwise a priori estimate:
\[
    \dist(\vp,\vp_j) \leq R_j,
\]
in $\Omega_j\bs\Sigma_j$ for $j\not\in J$,
where $R_j=1+T+T_j$ is clearly independent of the radius of
$\Ball$.  Using these estimates, we can now bound $\dist(\vp,\vp_j)$
on $\D\Omega_j$ for $j\in J$.
Therefore, for $j\in J$, the same argument, using the bounded subharmonic
functions $\nu_j = \sqrt{1 + \dist(\vp,\vp_j)^2}$ on
$\Omega_j\bs\Sigma_j$, yields $\dist(\vp,\vp_j) \leq R_j$,
for some $R_j$ independent of the radius of $\Ball$.
Setting $R=\max_j R_j$, Proposition~\ref{prop:uniform} follows.
\end{pf}

\subsection{Proof of the Main Theorem}

The proof of the uniqueness statement is almost the same as in
Proposition~\ref{prop:bounded}.  If $\vp$ and $\vp'$ are two solutions, then
they are asymptotic near each $\Sigma_j$, and asymptotic at infinity, hence
$\dist(\vp,\vp')^2$ is a bounded subharmonic function on
$\R^n\bs\Sigma$.  From
Lemma~\ref{lemma:max}, it follows that for any ball $\Ball\subset\R^n$
of radius $R$ centered at the origin:
\[
    \sup_{\Ball\bs\Sigma} \dist(\vp,\vp') \leq \sup_{\D\Ball\bs\Sigma}
    \dist(\vp,\vp'),
\]
Since the right hand side tends to zero as the radius $R$ of $\Ball$
tends to
$\infty$, it follows that $\dist(\vp,\vp')=0$, and hence $\vp=\vp'$.

To prove the existence of a solution,
we choose a sequence of radii $R_k\to\infty$,
let $\Ball_k$ be the ball of radius $R_k$ centered at the origin,
and denote by $\vph_k=(\uh_k,\vh_k)$
the solutions given by Proposition~\ref{prop:bounded} on $\Ball_k$.
We will use the uniform pointwise a priori bounds given in
Proposition~\ref{prop:uniform} to prove the convergence of
$\vph_k$ to a solution on $\R^n\bs\Sigma$.

We use the open cover
$\{\Omega'_j,\Omegat'_j\}$ introduced in section~\ref{sec:uniform}.
For any ball $\Ball$, define the space $\H'(\Ball)$ as the space of maps
$\vp\colon\Ball\bs\Sigma\to\Hy$ satisfying
\[
    \begin{cases}
    u-\uo \in H_1(\Ball); \\
    v-\vt_0 \in \bigcap_{j=1}^{N+1} H_1^{\vp_j}(\Ball;\R^{m-1}); \\
    \dist(\vp,\vp_j)\in L^\infty(\Omega'_j\bs\Sigma_j), \quad
    \forall j=1,\dots,N+1,
    \end{cases}
\]
and for $R>0$
denote by $\Hr'(\Ball)$ the space of maps $\vp\in\H'(\Ball)$ satisfying:
\[
    \begin{cases}
    \dist(\vp,\vpt) \leq R, &\text{in $\Omegat$}, \\
    \dist(\vp,\vp_j) \leq R, &\text{in $\Omega_j\bs\Sigma_j$, for
    $j=1,\dots, N+1$}.
    \end{cases}
\]

Now, fix $\Ball=\Ball_{k_0}$, let $\Ball'=\Ball_{R_0-1}$, and
consider the sequence of maps $\vph_k|_{\Ball}$ for $k\geq k_0$.
By Proposition~\ref{prop:uniform}, there is $R>0$ such that
$\vph_k\in\Hr'(\Ball)$. It follows that on
$\bigcup_{j=1}^{N+1}\Omega_j\bs\Sigma_j$, we have
$\norm{\uh_k-u_0}\leq R$.  Similarly, on $\Omegat$ we have
$\norm{\uh_k-\uo}\leq \norm{\uh_k-\ut} + \norm{\ut-\uo} \leq R'$
where $R'=R+\sup_{{\Omegat}} \norm{\ut-\uo}$.  Thus, we conclude
that $\norm{\uh_k-\uo} \leq C$ on $\Ball\bs\Sigma$, where
$C=\max\{R,R'\}$.  Furthermore, we have $\Delta(\uh_k-\uo)\geq0$
on $\Ball\bs\Sigma$.  Now the argument in the proof of
Lemma~\ref{lemma:max} shows that there is a constant $C'$
independent of $k$ such that
\[
    \int_{\Ball'} \norm{\nabla(\uh_k-\uo)}^2 \leq C',
\]
and similarly, using
$Q_{\vph_k}(\nabla\vh_k) \leq (1/2) \Delta(\uh_k-\uo)$, we deduce,
as in the argument leading to~\eqref{eq:vpt}, that
\[
    \int_{\Ball'} Q_{\vph_k}(\nabla\vh_k) \leq C''.
\]
Thus, we have a uniform bound $F(\vph_k) \leq C'''$.
It is straightforward
now to show that there is a subsequence, without loss of generality
$\vph_{k}$, which converges weakly
in $\H'(\Ball')$ and pointwise a.e.\ in $\Ball'$,
see~\cite[Proof of Proposition~1]{weinstein95}.

Repeating this argument for each
$k_0$, and using a diagonal sequence, it is
clear that we can choose a subsequence,
without loss of generality $\vph_{k}$ again, which converges
pointwise a.e.\ in $\R^n\bs\Sigma$, and
which for each ball $\Ball\subset\R^n$, converges
weakly in $\H'(\Ball)$ to a map $\vp\in\Hr'(\Ball)$.
For each open set $O\subset\subset\R^n\bs\Sigma$,
$\vph_k|_O$ is a
family of smooth harmonic maps with uniformly bounded energy into
$\Hy$, which maps into a fixed
compact subset of $\Hy$.  Hence, using
standard harmonic map theory, one can obtain uniform a priori
bounds in $C^{2,\alpha}(O)$ for $\vph_k$, and hence
we deduce that a subsequence converges
uniformly in $O$ together with two of its derivatives.
We conclude that $\vp$ is a harmonic map.
It remains to see that $\vp$ is asymptotic to $\vpt$ at infinity.
However, from the estimate~\eqref{eq:nubound} on $\nu$ which holds for each
$\vph_k$, we deduce that the same holds of $\vp$:
\[
    \dist(\vp,\vpt)\leq \sqrt{\Uo\bigl(2+\Uo\bigr)}
\]
Since $\Uo(x)\to0$ as $x\to\infty$, this
shows that $\dist(\vp,\vpt)\to0$ as $x\to\infty$ in
$\Ball\bs\Sigma$.  This completes the proof of the Main Theorem.

\section{$N$- Black Hole Solutions of the Einstein/Maxwell Equations}
\label{sec:einstein}

In this section, starting from a solution
$\vp\colon\R^3\bs\Sigma\to\HC$ of the Reduced Problem, we construct
a solution $(M,g,F)$ of the Einstein/Maxwell
Equations.

\subsection{The Spacetime Metric $g$}   \label{sec:g}

Let $\vp\colon\R^3\bs\Sigma\to\HC$ be a solution of the Reduced Problem.
Then $\vp=(u,v,\chi,\psi)$ satisfies on $\R^3\bs\Sigma$
the system~\eqref{eq:hu}--\eqref{eq:hp} of partial differential
equations, where however now, the Laplacian,
the divergence, the norms, and the inner-product are all with
respect to the Euclidean metric of $\R^3$.

Introduce cylindrical coordinates $(\rho,\phi,z)$ in $\R^3$, and
denote by $\xi$ and $\zeta$ the vector fields $\D/\D\phi$ and
$\D/\D z$ respectively.  Let $\ast$ be the Hodge star operator
of the Euclidean metric in $\R^3$ and
$\star= \rho^{-1}i_\xi\ast$ the Hodge star operator
of the flat metric $ds^2 = d\rho^2 + dz^2$ in the $\rho z$-plane.
Setting $\omega=2\, (dv + \chi\, d\psi - \psi\, d\chi)$
it follows from~\eqref{eq:hv} that the two-form $e^{4u} \ast
\omega$ is closed on $\R^3\bs\Sigma$, and hence, since $\omega$ is invariant
under $\xi$, the one-form
$e^{4u} i_\xi \ast \omega$ is closed on
$\R^3\bs\Sigma$.  Since $\R^3\bs\Sigma$ is simply connected, there is a
function $w$, defined up to an additive constant such that
\begin{equation}    \label{eq:gradw}
    dw = e^{4u} i_\xi \ast \omega = \rho \, e^{4u} \star \omega,
\end{equation}
see~\eqref{eq:dw}.

Now let $T_\vp$ be the stress tensor of the map $\vp$:
\[
    T_\vp(X,Y)
    = \inner{d\vp(X)}{d\vp(Y)} - \frac12 \norm{d\vp}^2 X \cdot Y,
\]
for $X,Y\in\R^3$, where $\inner{\cdot}{\cdot}$ denotes the inner product
in $\HC$.  Clearly $T_\vp$ is symmetric,
and since $\vp$ is a harmonic map $T_\vp$ is also
divergence free.  Thus, since $\zeta$ is a Killing field in $\R^3$,
$i_\zeta T_\vp$ is a divergence free vector field on
$\R^3\bs\Sigma$.  As before, it follows that
$i_\xi \ast i_\zeta T_\vp$ is a
closed one-form on $\R^3\bs\Sigma$, hence there is a function $\lambda$,
defined up to an additive constant such that
\[
    d(\lambda - u) = -2\,  i_\xi \ast i_\zeta T_\vp.
\]
Note that
\begin{multline*}
    -2 i_\xi \ast i_\zeta T_\vp =
    \rho \left[ \left( u_\rho^2 - u_z^2 + \frac{1}{4}
    e^{4u} (\omega_\rho^2 -
    \omega_z^2) + e^{2u} (\chi_\rho^2 - \chi_z^2 + \psi_\rho^2 -
    \psi_z^2)\! \right) d\rho \right. \\
    \left. {} + 2 \,  \left( u_\rho u_z
    + \frac{1}{4} e^{4u} \omega_\rho \, \omega_z +
    e^{2u} ( \chi_\rho \chi_z + \psi_\rho \psi_z ) \right)
    dz \right],
\end{multline*}
and compare with Equations~\eqref{eq:dlambda}.
It is clear that $w$ and $\lambda$ defined in this way are axially symmetric.
Furthermore, we deduce from~\eqref{eq:gradw}
that $e^{4u} \ast \omega = d\phi \wedge dw$.  Thus,
Equation~\eqref{eq:hp} implies that the two-form
$e^{2u} \ast d\psi + w \, d\chi \wedge d\phi$ is closed.  It follows that
the one-form $e^{2u} i_\xi \ast d\psi - w \, d\chi$ is closed, and hence
there is a function $\tilde\chi$ such that
\[
    d \tilde\chi = e^{2u} i_\xi \ast d\psi - w \, d\chi.
\]

Let $\A\subset\R^3$ be
the $z$-axis and let $g$ be the metric
on $M'=\R\times(\R^3\bs\A)$ given by the line element:
\begin{equation}    \label{eq:metric}
    ds^2 =
    - \rho^2 e^{2u} \, dt^2
    + e^{-2u} \bigl( d\phi - w\, dt \bigr)^2
    + e^{2\lambda} (d\rho^2 + dz^2).
\end{equation}
Define the one-form $A$ on $M'$ by:
\begin{equation}    \label{eq:A}
    A = -(\chi\, d\phi  + \tilde\chi\, dt),
\end{equation}
and let
\[
    F = dA = dt \wedge d\tilde\chi + d\phi \wedge d\chi.
\]
We have that $i_\xi F=d\chi$, and $i_\tau F =
d\tilde\chi$, where $\tau=\D/\D t$.
Clearly $\xi$ and $\tau$ are
commuting Killing fields of $(M',g)$ which
generate an abelian two-parameter group
$G=\R\times SO(2)$ of isometries leaving
$F$ invariant with timelike two-dimensional orbits, and hence $(M',g,F)$ is
stationary and axially symmetric.
Furthermore, it is not
difficult to check that $\omega$ is the twist of $\xi$, and that
$i_\xi \ast F = d\psi$ where here $\ast$ denotes the Hodge star operator of
the metric $g$.  It is then quite
straightforward to verify that $(M',g,F)$ satisfies the Einstein/Maxwell
equations.  Finally, since the $4N-1$ parameters
$\mu_j$, $r_j$, $\lambda_j$, and $q_j$ are
geometric invariants, these solutions form a $(4N-1)$-parameter family of
stationary and axially symmetric solutions of the Einstein/Maxwell
Equations.

\subsection{Conclusion}

We have shown the existence a $(4N-1)$-parameter family of
solutions of the Einstein/Maxwell Equations which are stationary and
axially symmetric, and which should be interpreted as $N$ co-axially
rotating charged black holes in equilibrium possibly
held apart by singular struts.  In order to complete this
interpretation, it is necessary to show that the metric $g$ and the
field $F$ can be extended smoothly across the axis of symmetry $\Sigma$.
This requires first the smoothness of the metric
and field components across $\Sigma$, which would follow from the smoothness
of $e^u$, $v$, $\chi$ and $\psi$ on $\Sigma\bs\D\Sigma$, where
$\D\Sigma$ denotes the set of $2N$ endpoints of $\Sigma$;
less regularity is expected on $\D\Sigma$, see~\cite{weinstein92}.
For the target $\Hy=\HR$ corresponding to the Einstein/Vacuum Equations
this was established in~\cite{weinstein90}.
Regularity of harmonic maps with prescribed singularities
into $\HR$ was studied in further generality in~\cite{litian92}.

Even after smoothness is established, there is still the possibility of a
conical singularity on the bounded components of $\Sigma$.  As in the vacuum
case, this is to be interpreted as a singular strut holding the black holes
apart, and the angle deficiency can be related to the force between these.
Finally, to prove that $g$ is asymptotically flat, the decay estimate
obtained at infinity must be sharpened, see~\cite{weinstein92}.

These questions will be addressed in a future paper.

\end{document}